%% file: main.tex
\PassOptionsToPackage{prologue,dvipsnames,table}{xcolor}

\documentclass[acmtog,screen, nonacm]{acmart}
\usepackage{tikz}
\usepackage{xcolor}

\usetikzlibrary{shapes, arrows.meta, decorations.markings}
\usepackage{float}
\usepackage{caption}
\usepackage{booktabs} 
\usepackage{bbm}
\usepackage{url}
\usepackage{enumitem}
\usepackage{booktabs}
\usepackage{siunitx}
\usepackage{xspace} 
\usepackage{nicefrac}
\usepackage{algorithm}
\usepackage{algpseudocode}
\usepackage{empheq} 
\usepackage{mathtools}
\usepackage{algpseudocode}
\usepackage{amsmath}
\usepackage{acronym}
\usepackage{abbrv}

\definecolor{codegreen}{rgb}{0,0.6,0}
\definecolor{codegray}{rgb}{0.5,0.5,0.5}
\definecolor{codepurple}{rgb}{0.58,0,0.82}
\definecolor{backcolour}{rgb}{0.96,0.96,0.98}
\usepackage{listings}

\setcitestyle{square}

\title{Lightweight Predictive 3D Gaussian Splats}

\author{Junli Cao}
\email{jcao2@snapchat.com}
\affiliation{
  \institution{Snap Inc.}
  \city{Los Angeles}
  \country{USA}
}

\author{Vidit Goel}
\email{vgoel@snapchat.com}
\affiliation{
  \institution{Snap Inc.}
  \city{Los Angeles}
  \country{USA}
}

\author{Chaoyang Wang}
\email{cwang9@snapchat.com}
\affiliation{
  \institution{Snap Inc.}
  \city{Los Angeles}
  \country{USA}
}
\author{Anil Kag}
\email{akag@snapchat.com}
\affiliation{
  \institution{Snap Inc.}
  \city{Los Angeles}
  \country{USA}
}

\author{Ju Hu}
\email{jhu3@snapchat.com}
\affiliation{
  \institution{Snap Inc.}
  \city{Los Angeles}
  \country{USA}
}

\author{Sergei Korolev}
\email{skorolev@snapchat.com}
\affiliation{
  \institution{Snap Inc.}
  \city{Los Angeles}
  \country{USA}
}

\author{Chenfanfu Jiang}
\email{cffjiang@math.ucla.edu}
\affiliation{
  \institution{University of California, Los Angeles}
  \city{Los Angeles}
  \country{USA}
}

\author{Sergey Tulyakov}
\email{stulyakov@snapchat.com}
\affiliation{
  \institution{Snap Inc.}
  \city{Los Angeles}
  \country{USA}
}

\author{Jian Ren}
\email{jren@snapchat.com}
\affiliation{
  \institution{Snap Inc.}
  \city{Los Angeles}
  \country{USA}
}

\usepackage{algorithm}
\usepackage{amsmath}

\usepackage{algorithmicx}
\usepackage{algpseudocode}
\usepackage{multirow}
\usepackage{caption}
\usepackage{subcaption,bm}
\usepackage{stmaryrd}
\usepackage{wrapfig}
\algdef{SE}[DOWHILE]{Do}{doWhile}{\algorithmicdo}[1]{\algorithmicwhile\ #1}

\AtBeginDocument{%
  \providecommand\BibTeX{{%
    \normalfont B\kern-0.5em{\scshape i\kern-0.25em b}\kern-0.8em\TeX}}}

\begin{document}

\input{sec/0_abs}

\begin{teaserfigure}
\includegraphics[width=\linewidth]{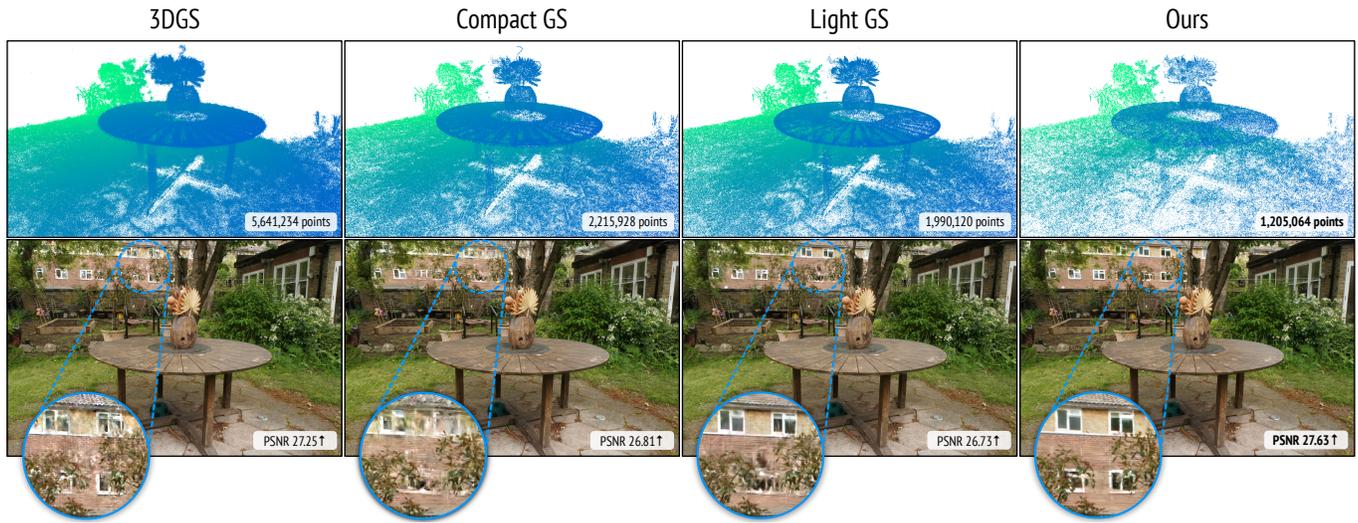}
\caption{\emph{Top}: We show point clouds of the \texttt{Garden} scene~\cite{mip-nerf360} obtained using different methods. Our method features the smallest number of points to store. 
\emph{Bottom}: Images rendered using the compared methods. Ours shows the best PSNR. We magnified a region highlighted with blue. Note, that despite significantly smaller storage requirements, our method has the highest fidelity, being able to best reconstruct the detailed structure of the image. \emph{Zoom-in for greater detail.}
}
\label{fig:teaser}
\end{teaserfigure}

\maketitle
\input{sec/1_intro}

\input{sec/2_related_work}

\input{sec/3_method}
\input{sec/4_experiments}
\input{sec/5_conclusion}

\bibliographystyle{ACM-Reference-Format}
\bibliography{main}
\clearpage
\appendix
\input{sec/supp}
\end{document}

%% file: sec/0_abs.tex
\begin{abstract}

Recent approaches representing 3D objects and scenes using Gaussian splats show increased rendering speed across a variety of platforms and devices. 
While rendering such representations is indeed extremely efficient, storing and transmitting them is often prohibitively expensive. 
To represent large-scale scenes, one often needs to store millions of 3D Gaussians, occupying gigabytes of disk space. 
This poses a very practical limitation, prohibiting widespread adoption.
Several solutions have been proposed to strike a balance between disk size and rendering quality, noticeably reducing the visual quality.
In this work, we propose a new representation that dramatically reduces the hard drive footprint while featuring similar or improved quality when compared to the standard 3D Gaussian splats. When compared to other compact solutions, ours offers higher quality renderings with significantly reduced storage, being able to efficiently run on a mobile device in real-time~\footnote{More demo examples in our webpage:~\url{https://github.com/plumpuddings/LPGS}.} 
Our key observation is that nearby points in the scene can share similar representations. 
Hence, only a small ratio of 3D points needs to be stored. 
We introduce an approach to identify such points—called \emph{parent} points. The discarded points—\emph{children} points—along with attributes can be efficiently predicted by tiny MLPs. 

\end{abstract}

%% file: sec/1_intro.tex
\section{Introduction}

Gaussian Splatting (3DGS)-based methods are taking the graphics and vision communities by a storm~\cite{luiten2023dynamic,wu20234d,yang2023deformable}. They promise to strike the right balance between high-fidelity rendering, fast convergence, and efficient inference~\cite{3dgs}. The latter two benefits make 3DGS-based methods superior to Neural Radiance Fields (NeRFs)-based techniques~\cite{nerf,nerf-in-the-wild,mipnerf}. Indeed, while NeRFs~\cite{mip-nerf360} show high-fidelity renderings too, apart from several exceptions~\cite{mobiler2l,r2l,mobilenerf, muller2022instant}, their training and inference time is often prohibiting real-time and edge-based applications. 

3DGS-based approaches represent a 3D scene using an explicit, point-based representation~\cite{aliev2020neural}. Geometry and color are stored as millions of 3D Gaussians. The 3D Gaussians are efficiently rasterized to 2D images, with much faster rendering than neural volumetric rendering approaches~\cite{3dgs}. However, to represent sophisticated geometry and texture, especially for large-scale scenes, a significant amount of points along with their attributes need to be stored, amounting to gigabytes of storage. 

In a world of connected devices, real-time experiences and applications, this storage requirement imposes a heavy toll on the hard-drive and the transmission bandwidth. Hence, several initial solutions have been proposed to reduce the storage for 3DGS, such as incorporating a sparse voxel grid~\cite{scaffoldgs} or applying more aggressive pruning of the 3D points~\cite{lightgs,compactgs}.
Yet, existing studies still suffer either from large storage requirements~\cite{scaffoldgs} or inferior rendering quality compared to 3DGS~\cite{lightgs, compactgs}.

In this work, we introduce a lightweight Gaussian Splat representation, with storage significantly reduced, and featuring superior rendering quality. Empowered with our approach, a practitioner does not have to compromise the quality. The key to our approach is a realization that not all points are equally important for rendering (also discussed in~\cite{scaffoldgs}). In fact, only a fraction of points need to be stored. 
We refer to such points as \emph{parent} points. The remaining points, named as \emph{children} points, do not have to be stored, but can be \emph{predicted} instead. 
Besides, we show that even the Gaussian attributes can be estimated during rendering. 

Fig.~\ref{fig:teaser} shows the Garden scene~\cite{mip-nerf360} reconstructed by the standard Gaussian Splats~\cite{3dgs}, Compact GS~\cite{compactgs}, Light GS~\cite{lightgs} and the proposed approach. First, we observe a significantly reduced density of points in the point cloud reconstructed by our approach. This, and the predicting of the attributes instead of storing them, significantly reduces the storage requirement for our method. Second, we show improved PSNR scores and visual quality, when we zoom-in into the details of the rendered images.

To identify the \emph{parent} points, we first allow every point to be considered as a \emph{parent} point, and be used to  predict its \emph{children} points. 
We allow children nodes to be promoted to new parents when required.
If found unnecessary during optimization, we can delete \emph{parent} points together with all their \emph{children} points. Intuitively, this allows certain regions with sophisticated geometry to contain more points for accurate modelling. Therefore, starting with points obtained using SfM~\cite{sfm}, our method learns a compact set of \emph{parent} points. \emph{Children} points do not have to be stored and can be predicted instead. We use a 3D hash grid~\cite{ngp} to encode the offsets used to estimate the 3D locations of such points. Besides that, in the same hash grid, we store the features necessary for predicting Gaussian attributes for each point. Intuitively, nearby points represent similar geometry and texture and, thus, could be used when predicting the attributes. To do so, we introduce an attention-based mechanism attending to both parent and child points. Prediction is performed using shallow 2-layer fully-connected networks, adding a negligible overhead to the process. 

Equipped with the above contributions, to represent large-scale real-world scenes, our method requires dramatically less storage, while having comparable or improved quality metrics. Compared with the original 3DGS~\cite{3dgs}, we show up to $20\times$ reduction on average in hard-drive footprint, with improved PSNR and comparable SSIM and LPIPS. Compared with current voxel grid-based approach~\cite{scaffoldgs}, our method provides $5\times$ storage reduction, while showing improved metrics on the dataset introduced in~\cite{mip-nerf360}, and $2\times$ reduction and comparable metrics on Tanks\&Temples~\cite{tank} (see Tab.~\ref{tab:exp}). Besides, our method offers a reader a practical means to satisfy the storage requirements of their task at a hand. In Fig.~\ref{fig:size-plot} we show several configurations depending on the total hard-drive footprint. Our smallest configuration, which is 2$\times$ smaller than the smallest prior work~\cite{lightgs}, shows similar scores. Our largest model, which is still smaller than all other works, reaches higher quality. Due to smaller storage requirements, efficient rendering, and superior quality our method can unlock wide adoption of GS-based applications on resource-constrained devices, such as mobiles.

%% file: sec/2_related_work.tex
\section{Related Work}

\noindent\textbf{Novel View Synthesis.}
Research on rendering scenes from unseen viewpoints with photorealism has evolved over several decades~\cite{greene1986environment,chen2023view,levoy2023light,buehler2023unstructured,srinivasan2019pushing}. Traditional approaches typically rely on explicit depth estimation to warp pixels for generating novel views~\cite{kalantari2016learning,penner2017soft,choi2019extreme,riegler2021stable}. However, the accuracy of depth estimation algorithms is critical, and handling disocclusions during rendering adds complexity. An alternative approach involves Multiplane Images (MPI)~\cite{zhou2018stereo,srinivasan2019pushing,flynn2019deepview}, which learn a representation associating objects within the scene with fronto-parallel layers. This structured representation facilitates efficient rendering from different viewpoints while preserving depth relationships and occlusions. More recently, Neural Radiance Fields (NeRF)~\cite{nerf} have gained popularity for their ability to achieve highly realistic rendering, even in scenarios involving complex view-dependent lighting effects such as transparency and reflectance. However, the weakness of NeRF lies in its volumetric rendering formulation, which necessitates sampling a large number of points per ray to render a single pixel. This high computational cost limits the usage of NeRF for real-time or on-device applications. While efforts to reduce computational requirements for volumetric rendering have been a focus of recent research~\cite{liu2020neural,neff2021donerf,garbin2021fastnerf,reiser2021kilonerf,lindell2021autoint,yu2021plenoctrees,muller2022instant,fridovich2022plenoxels,lombardi2021mixture,mobiler2l,lightspeed}, point-based rendering, particularly 3D Gaussian Splatting (3DGS)~\cite{3dgs}, 
presents another promising direction for real-time view synthesis.

\noindent\textbf{Point-based Rendering.}
The basic form of point-based rendering rasterizes an unstructured set of points and utilizes graphics APIs for rendering~\cite{botsch2005high,ren2002object}, providing speed and editing flexibility, yet often resulting in artifacts such as holes and outliers. To address this, researchers have explored learnable point-based representations. One approach is to associate neural features with each point, rendering images using CNNs with rasterized feature maps as input~\cite{Rakhimov_2022_CVPR,aliev2020neural,kopanas2021point,feng2022neural}. Another relevant approach is to learn parameters of point primitives such as ellipsoids, or surfels~\cite{wiles2020synsin,yifan2019differentiable,gross2011point,insafutdinov2018unsupervised,lin2018learning,3dgs}. Notably, 3D Gaussian Splatting~\cite{3dgs} learns anisotropic 3D Gaussians, treating them as volumes of radiance and splatting them to 2D images using alpha-blending. This technique achieves higher-quality results compared to NeRF-based methods and can render high-resolution images in real-time. One drawback of 3D Gaussian splatting and many point-based methods is their inefficiency in terms of storage complexity. This is due to their unstructured representation, which ignores spatial correlations between points. Consequently, they require storing a large number of points to represent complex scenes, often numbering in the millions. Therefore, reducing the consumption of storage in 3D Gaussian splatting has become an emerging research topic.

\noindent\textbf{Efficient Representation for 3D Gaussian Splatting.}
There are two main directions to improve the efficiency of 3D Gaussian splatting representations: reducing the number of points and compressing storage for point-wise attributes (\eg, spherical harmonics and geometric properties). We find three concurrent works which are most related to ours. {LightGS}~\cite{lightgs} introduces a  point pruning and recovery process to minimize redundancy in Gaussian splats, utilizes distillation and pseudo-view augmentation to distill spherical harmonics to a lower degree, and employs quantization to further reduce storage. While LightGS achieves considerable storage reduction, it results in noticeable fidelity degradation compared to the original Gaussian splatting due to quantization. {CompactGS}~\cite{compactgs} proposes using a grid-based neural field to implicitly represent view-dependent colors rather than explicitly storing spherical harmonics per point, offering promising storage efficiency without significant fidelity loss. {ScaffoldGS}~\cite{scaffoldgs} suggests distributing local  splats using anchor points, re-parameterizing point positions relative to these anchors to enable anchor-based point growing and pruning strategies for redundancy reduction in 3DGS.
Our method shares similarities with CompactGS~\cite{compactgs} and ScaffoldGS~\cite{scaffoldgs} while exhibits crucial \emph{differences}. \emph{First}, unlike CompactGS, we utilize a combination of neural fields and self-attention layers to predict not only view-dependent colors but also geometric properties. \emph{Second}, in contrast to both approaches which explicitly store the position of every point in the point cloud, our method only stores a small subset of points, referred to as \emph{parent} points, while predicting the remaining points on-the-fly during rendering. This substantially reduces memory footprint. \emph{Third}, we introduce a new point growing and pruning strategy that allows our approach to achieve high-fidelity rendering with fewer points.

%% file: sec/3_method.tex
\input{figures/pipeline}

\section{Preliminaries}
3D Gaussian Splatting (3DGS)~\cite{3dgs} represents a scene with 3D points $\textbf{x}$. The points are initialized with a coarse point cloud obtained using Structure-from-Motion (SfM)~\cite{sfm}.
These Gaussians, $G(\textbf{x})$, serve as the anisotropic volumetric splats defined by their position (mean $\mu$) and 3D covariance ($\Sigma$) as 
$G\left(\textbf{x}\right) = e^{-\frac{1}{2}\left( \textbf{x} - \mu\right)^T \Sigma^{-1}\left( \textbf{x} - \mu\right)}$.

To ensure $\Sigma$ remaining positive semi-definite during optimization, it is represented with an equivalent yet effective formulation with the scaling matrix $S$ and the rotation matrix $R$, such that $\Sigma = RSS^TR^T$.
The attributes of the 3D splats (\eg, location, covariance, and opacity) together with the directional appearance of the radiance filed, represented via the spherical harmonics (SH)~\cite{plenoxels}, are end-to-end learned using optimization. During the optimization the number of Gaussians is changed via cloning, splitting, and pruning operations. 

To render an image, 3D $G(\textbf{x})$ are first transformed into 2D Gaussians (denoted as $G'(\textbf{x})$)~\cite{ewa}. 
3DGS uses an efficient tiled-based rasterizer that presorts primitives for the entire image, allowing fast $\alpha$-blending of anisotropic splats. The color $C$ of a pixel is computed by blending $N$ 2D Gaussians that overlap at the pixel as: $C = \sum_{i\in N}c_i \alpha_i G'_i(\textbf{x})\prod_{j=1}^{i-1}(1 - \alpha_j G'_j(\textbf{x})),$
where $c_i$ represents view-dependent colors for each splat, $\alpha_i$ is the opacity.
Thanks to the highly optimized rasterizer for modern GPUs, 3DGS render high-fidelity scenes in real-time across a number of platforms. 

These benefits come with a cost. 3DGS require a significant number of 3D Gaussians, sometimes needing gigabytes for complex large-scale scenes. This requirement limits their application on edge devices, as downloading gigabytes over the network and storing them is hardly feasible or practical for edge devices at scale. 
\section{Method}
We show a high-level overview of our approach in Fig.~\ref{fig:pipeline}. Our key motivation is that particular splats carry greater importance than others, due to their position, opacity, scale, \etc~\cite{scaffoldgs}. We show that these important splats can be used to derive the attributes---position, color, scale, \etc---of other splats using a small neural network. This allows us to store only the important splats along with the weights of the neural network. To do so, we represent a scene using a forest of a depth-$1$ tree structures, where we use \emph{parent} nodes to represent the important splats using which we predict $K$ \emph{children} nodes. 
Formally, we represent a scene using $\mathcal{S} = \{\mathcal{X}_1 , \mathcal{X}_2, \dots \mathcal{X}_n\}$, where $\mathcal{X}_i$ is tree and each node contains the attributes, such as position ($x$), color ($c$), opacity ($\alpha$), scale ($s$), and rotation ($r$). This representation can be stored very efficiently, as for each tree we need to save only the positions and scales of \emph{parent} nodes and small neural network shared across the trees, to predict all the other attributes of the tree.

\subsection{Neural Representation for Lightweight Predictive Splats} \label{sec:pipeline}
We design the architecture to model the close relationship between a parent and children nodes. In particular, we assume that the children nodes are in the vicinity of parent node and have similar physical attributes such as shape, color and opacity. We satisfy these requirements by using a hash-grid based approach~\cite{ngp,chen2023dictionary} as our representation. Hash-grids have an inherent property to return similar features when queried with the points located nearby. Below we describe in detail how a tree ($\mathcal{X}_i$) can be represented in storage efficient manner. In what follows we drop the index $i$.

Given a hash-grid $\mathcal{H}(\cdot)$ shared across the trees and parent node positions $x_p$, we query the features as $f = \mathcal{H}({x}_p)$  and use them to predict all the attributes of the tree. Specifically, we divide $f$ into two halves $f\equiv\{f_\Delta\in \mathbb {R}^{D/2}, f_a\in \mathbb {R}^{D/2}\}$, where the first half ($f_\Delta$) represents displacement and is used to predict the position of children. The second half ($f_a$) is used to predict rest of the attributes. 

\noindent \textbf{Predicting Position.} 
We want children and parent nodes to represent similar geometry and appearance. Hence, children should be located in the vicinity of the parents nodes. We model the position of children as their displacement from their parent nodes. For the parent we predict the position of $k^\mathrm{th}$ child using $ x_k = x_p + g_\mathrm{pos}(f_\mathrm{\Delta})[k]$ where $g_{\mathrm{pos}}$ is an MLP with output shape $K\times3$.

Having the positions of all nodes in the tree, we can predict the rest of the attributes, such as scale, rotation, color, and opacity. We reuse the hash-grid to get the attribute feature (${f_\mathrm{a}}_k$) for $k^\mathrm{th}$ child node using $\mathcal{H}(x_k)$. A naive approach to extract the remaining attributes using $f_\mathrm{a}$ and ${f_\mathrm{a}}_k$ is to pass the latter to an MLP get scale, rotation, color and opacity. We found such approach to be sub optimal. A hash-grid representation implicitly makes the representation of spatially points similar. There is no mechanism to share information between the features after they are computed. Since there is relation between physical attributes of the parent and children nodes, having such information sharing mechanism is beneficial. To this end, we propose a modified self-attention mechanism to better capture the inter-dependencies between children and parent nodes. To do so, we first obtain the aggregated feature $\mathcal{F}_\mathrm{a}$ $\in \mathbb{R}^{K+1 \times D/2}$ by concatenating features of all the nodes in the tree, such as $\mathcal{F}_\mathrm{a} = \texttt{Concat}(\{f_\mathrm{a}, ({f_\mathrm{a}}_1, \dots, {f_\mathrm{a}}_K \})$, where $\texttt{Concat}$ is a concatenation operation. We then apply a modified self-attention operation on $\mathcal{F}_{\mathrm{a}}$ to get the final feature $\mathcal{F}^\prime_{\mathrm{a}}$:
\begin{equation}\label{eqn:fusion}
\begin{aligned}
    \mathcal{F}^\prime_\mathrm{a} = \mathcal{F}_\mathrm{a} + \lambda \sigma(\frac{\mathcal{P}_1(\mathcal{F}_\mathrm{a})* \mathcal{P}_2(\mathcal{F}_\mathrm{a})^T}{\sqrt{d}})*\mathcal{F}_\mathrm{a},
    \end{aligned}
\end{equation}
where $\sigma(\cdot)$ is a Softmax function, $\mathcal{P}_i(\cdot)$ is a projection matrix, $d$ is a scaling factor set as $D/2$, $\lambda$ is a hyper-parameter for balancing the information trade-off from the attention mechanism and $*$ denotes the matrix multiplication. In contrast to vanilla attention mechanism~\cite{attn}, we do not apply positional embedding, so that Eq.~\ref{eqn:fusion} is permutation invariant which is an important property to maintain while working with point clouds~\cite{pointnet}.
Further, we use the unprojected $\mathcal{F}_\mathrm{a}$ when multiplying with $\sigma(\cdot)$, since we  empirically found no performance gain by projecting $\mathcal{F}_\mathrm{a}$. 
Next, we split $\mathcal{F}^\prime_\mathrm{a}$ in $K + 1$ attribute feature vector to predict the remaining attributes for each node in the tree. 

\noindent\textbf{Predicting Scale and Rotation:} It is vital to properly initialize the scale of Gaussians for stable training. For instance, Gaussians with small scales make minimal contributions to the rendering quality, mainly because of their limited volume. 
In contrast, large Gaussians can potentially contribute to every pixel during rasterization, leading to a significant amount of GPU memory. Hence, to make training stable and minimize storage needs at the same time, we adopt a middle-ground strategy. More specifically, we represent the scales of children as a scaled version of their parents ($s_\mathrm{p}$): $s_k = \hat{s}_k \ s_p$ where $\hat{s}_k$ is predicted by an MLP. In case of rotation, we directly regress it for both parents and children nodes using the corresponding attribute feature vector.
We share the weights of the MLP to regress both scale and rotation. 
We experimentally found, that including position of node ($x_k$), the distance of the point to the center of the axis aligned bounding box ($b_k$) along with attribute feature ((${f^\prime_\mathrm{a}}_k$)  improves performance: $\hat{s}_k, r_k = g_{rs}({f^\prime_\mathrm{a}}_k, x_k, b_k)$.

\noindent\textbf{Predicting Color and Opacity:} 3DGS uses degree-$3$ spherical harmonics (SH) for view-dependent color representation~\cite{3dgs}. However, we find such design is unnecessary and the color can be directly predicted using from feature vectors and a viewing direction. We use an MLP that takes them as an input and directly predicts the color as output, $c_k = g_c({f^\prime_\mathrm{a}}_k, d_k)$ where $d_k$ is the viewing direction of the node in the tree. This helps in reducing the storage by a significant amount as previously each splat had to store the spherical harmonics individually. 
To predict opacity, we use another MLP with inputs as ${f^\prime_\mathrm{a}}_k$ and the position of the node to get corresponding opacity, $o_k = g_o({f^\prime_\mathrm{a}}_k, x_k)$. 

We described all the operations above for a single tree. The same operation is extended for all the trees. Further, the neural networks for all the operations share their weights across all the trees.
To summarize, the proposed representation can efficiently represent the tree structure using hash-grid based neural representations  $\mathcal{H}(\cdot)$ and a few MLPs. We only need to store position and scale of parent nodes and the weights of our neural networks, while the rest of the properties of the tree is regressed as described above.

\subsection{Adaptive Tree Manipulation}
\label{subsec:treemani}
3DGS  starts by using an initial point cloud from SfM~\cite{sfm}. To allow for some flexibility in the point cloud structure they propose several techniques to add and delete points during optimization. The techniques are based on position gradient, size, and opacity of the gaussians.
Extending these techniques to our depth-$1$ tree structure of  is not straightforward. Treating parent nodes as individual points and following 3DGS can lead to sub-optimal or incorrect results (see Tab. \ref{tab:core}). 
In the proposed depth-$1$ tree representation both parents and children impact the rendering quality. Hence, to allow for flexibility one needs to incorporate children nodes, when adding or deleting trees in the scene.

Along with maintaining the gradient of position of parent nodes we also track them for the children nodes. When the gradient of a child node is above a certain threshold $\tau^{c}_\mathrm{pos}$ we split the tree and promote the child node to become a new parent. This is crucial to represent complicated regions in the scene where there might not be many parent nodes. Once the child has been promoted to parent node we apply clone and split operations to all the parents following similar practices in 3DGS. This gives flexibility to children nodes to add trees and help in the densification of region if necessary. 

To delete an entire tree we only rely on the statistics of parent node. This is because if a child node was important then it would have been already promoted to become a new parent. Hence, we can safely delete the current parent that will in turn delete all the corresponding children nodes. Specifically, for deleting the trees we track the scale and opacity of the parent nodes and delete them if they are below a certain threshold similar to 3DGS.

\subsection{Training}
Our model, including the hash-grid and MLPs, is \emph{end-to-end} learnable. These components are guided by the $\mathcal{L}_1$ loss  between the rendered images and ground-truth images along with a D-SSIM loss, such that:
\begin{equation}\label{eqn:loss}
\begin{aligned}
  \mathcal{L} &= (1 - \beta)\mathcal{L}_1 + \beta \mathcal{L}_{\mathrm{D-SSIM}},
\end{aligned}
\end{equation}
where $\beta$ is set as $0.2$ following the setting in ~\cite{3dgs}. 

We use a warm-up training scheme that helps in convergence of the model~\cite{3dgs}. The warm-up consists of training the model in a low resolution setting, eventually moving to higher resolution after a certain number of steps have been completed. We found that the warm-up strategy is crucial to correctly position the splats and densify the regions.
Without the warm-up, the model struggles to populate enough splats in the background area, despite the importance of the area resulting in substandard performance. Please refer to  \emph{Supplementary} material for details.

%% file: figures/pipeline.tex
\begin{figure*}[!t]
\centering
\includegraphics[width=1\textwidth]{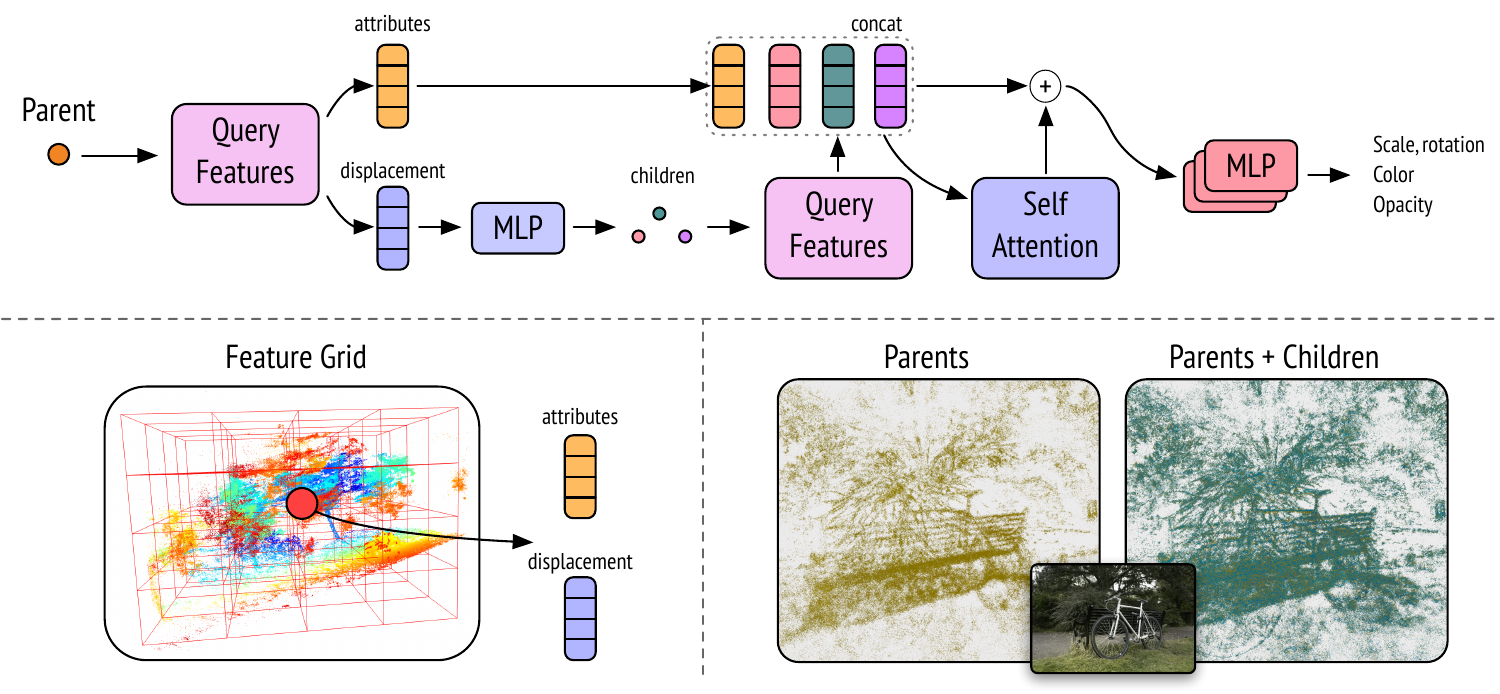}
\caption{\emph{Top}: We use a \emph{parent} node to estimate its \emph{children} nodes and the Gaussian attributes. The \emph{parent} node retrieves a pair of features, used for attributes $f_\mathrm{a}$ and displacement ($f_\mathrm{\Delta}$) prediction, from the feature grid. The displacement features $f_\mathrm{\Delta}$ are used to estimate the positions of \emph{children} nodes. To estimate the Gaussian attributes, such as scale, rotation, color, and opacity, attribute features $f_\mathrm{a}$ are aggregated with self-attention.
\emph{Bottom Left}: The process of querying the features. \emph{Bottom Right}: A visualization of \emph{parent} nodes and their predicted \emph{children} nodes.
}
\label{fig:pipeline}
\end{figure*}

%% file: sec/4_experiments.tex
\section{Experiments}\label{sec:exp}
Our aim is to build an efficient GS representation, with low storage requirements, high-fidelity rendering and real-time execution. These features are particularly important for resource constrained devices such as mobile phone, as transferring large volumes of data across the cell network and storing it locally is not practical. 

We provide extensive experiments to evaluate and validate the contributions introduced in the previous section. 

\noindent\textbf{Implementation Details}. 
 We use Instant-NGP~\cite{ngp} as our hash-grid owing to its compact and efficient design and 2 layers MLP for all the MLPs. Following the practices in~\cite{mip-nerf360, ngp} we use scene contraction to map the position into $[0, 1]$ before feeding it to the Instant-NGP. This helps bring the splats that are occasionally outside the Axis-Aligned-Bounding-Box (AABB) due to the densification of splatsalong with the position updates. We estimate the AABB with the initial COLMAP~\cite{colmap} point cloud. This algorithms of AABB estimation and contraction are included in \emph{supplementary}.
 We set $\lambda = 0.5$ for all the experiments and train the model for $30K$ steps where initial $7.5K$ steps is warm-up stage. The number of \emph{children} splats 
 ($K$) used in our experiment varies across scenes, with at most 2 children.  Please refer to \emph{Supplementary} material for details.

\input{Table/comparison_table_full}

\noindent\textbf{Dataset and Metrics} We evaluate our method using \emph{seven} scenes from the Mip-NeRF $360^\circ$ dataset~\cite{mipnerf}, \emph{two} scenes from Tank\&Temples~\cite{tank}, and \emph{two} scenes from Deep Blending~\cite{db}. We use the widely adopted metrics like PSNR, SSIM~\cite{ssim}, and LPIPS~\cite{lpips} to assess the quality for image reconstruction. We also report the storage size (in MB) for various methods along with their on-device capabilities. We benchmark the Gaussian Splatting based methods on iPhone 14 with our implementation of the mobile application. 
We report three configurations of our method named C1, C2, and C3 by varying feature dimension $D$ of the hash-grid $\mathcal{H}$. C1 is our smallest model with $D = 32$ followed by C2 with $D = 48$ and C3 is largest with $D = 64$.Since our framework adds and removes points during optimization, the final storage for each model can vary. For each dataset we report the average size of all the scenes within one configuration. The metrics, too, are averaged over all scenes of each dataset. Per-scene quantitative results are in the \emph{Supplementary} material.

\input{figures/size_plot}

\subsection{Comparison Results}

\input{figures/exp}
\noindent\textbf{Quality vs Storage}. First, we show that our approach provides a practical means of satisfying diverse technical requirements. We can reduce or increase the feature dimension of the hash-grid and the number of points, while still maintaining similar or superior rendering quality. In Fig.~\ref{fig:size-plot}, we plot PSNR, evaluated on the dataset introduced by~\cite{mip-nerf360}, for contemporary models as well as for three configurations of our approach. Our smallest configuration is almost 50\% smaller than the smallest prior work (LightGS~\cite{lightgs}), and shows the same rendering quality. Our largest configuration, which is still 32\% smaller than the smallest existing work, shows significantly increased PSNR. To give the reader a better perspective, we also plot conventional works with large hard-drive footprint~\cite{scaffoldgs,3dgs}. Our largest configuration, which uses only 20\% of ScaffoldGS~\cite{scaffoldgs} and only 4.5\% of 3DGS~\cite{3dgs} storage, shows higher quality than both of these much larger works. 
These advantages of our method are crucial for mobile deployment.
Less disk storage also helps in speeding up transmission that significantly impacts user experience when sharing content.

\input{figures/atm}
\noindent\textbf{Quantitative Results}.
Tab.~\ref{tab:exp} shows the quantitative results performed on real-world scenes, spanning from large-scale urban landscapes to intricate indoor and outdoor environments. 
We compare our approach against NeRF-based methods, 3DGS~\cite{3dgs}, and concurrent works (\ie, LightGS~\cite{lightgs}, CompactGS \cite{compactgs}, and ScaffoldGS \cite{scaffoldgs}). 
On the Mip-NeRF 360$^\circ$ dataset, our approach achieves the \emph{best} PSNR among all the approaches. Compared with 3DGS~\cite{3dgs}, we obtain a significant storage \emph{reduction}, \ie, $19.5\times$, and require $3.5\times$ \emph{fewer} 3D points.
On the Tank\&Temples~\cite{tank} dataset, although ScaffoldGS~\cite{scaffoldgs} has better PSNR than our approach, our storage is almost $2.4\times$ \emph{smaller} than ScaffoldGS.

Compared with 3DGS~\cite{3dgs} on this dataset, we require 
$1.9\times$ \emph{fewer} 3D points and $11.3\times$ \emph{less} storage. 
Lastly, on the Deep Blending~\cite{db} dataset, our method has \emph{higher} PSNR and $19\times$ storage \emph{reduction} than 3DGS~\cite{3dgs}.

\textit{It is worth noting that there always exist a configuration of our method where we achieve smallest size and best PSNR when compared with recent works~\cite{compactgs, lightgs} on all the datasets.}

\noindent\textbf{Qualitative Results}.
Fig.~\ref{fig:eval} demonstrates the high-quality rendering of our method produced using C3 configuration across $5$ example scenes covering all the datasets. We see various examples where our method outperforms previous compression works. We can see our models can better capture background details (row 3, 5), better capture reflections (row 2) while being the smallest or of comparable size. It can also capture intricate details where other methods fail such as ceilings (row 1, 4). We encourage the reader to check out the webpage for more results.

\noindent\textbf{On-Device Capability}. 
We explore the feasibility of running splatting based methods on mobile devices. We use iPhone14 as the platform to implement a testing application. For fair comparison, we unpack splats from all methods to a standard 3DGS format~\cite{3dgs} for rendering. We observe Out-of-Memory error when running all scenes from the three benchmark datasets for 3DGS and ScaffoldGS, owing to their large number of splats. Our method can successfully run on device, and achieves smaller and  better rendering quality compared to LightGS and CompactGS.

\input{Table/core_and_attn}
\subsection{Ablation Analysis}\label{sec:ablation}
We perform comprehensive analysis on various components of our methods using our C3 configuration. Here we choose two representative scenes to perform experiments: one unbounded outdoor scene \texttt{Bicycle} from Mip-NeRF 360$^\circ$ dataset~\cite{mip-nerf360} and one indoor scene \texttt{Playroom} from the Deep Blending dataset~\cite{db}. We report the best PSNR that is achieved within $10$K steps for all experiments. 
\input{Table/attn}
\input{Table/arch}

\input{figures/promoted_parents}

\noindent \textbf{Importance of Hash Grid.} We replace the hash grid with the frequency encoding of the 3D position followed by a $2$-layer MLP to output a $D=64$-dimensional feature vector, which has the same dimension as the one from hash grid. We denote the setting as \emph{FE}.
    Without hash-grid we see a significant drop in the performance, highlighting the importance of the feature alignment encoded within the spatial hash grid. 

\noindent \textbf{Importance of Attention Mechanism.}  It can be seen that when we remove self attention mechanism between the nodes of the tree it is detrimental to the performance (Tab.~\ref{tab:core} w/o Attn). This validates our motivation that there is relation between various physical attributes of the nodes of tree hence there needs to be a mechanism to facilitate the sharing of information. We also noticed that adding attention mechanism reduces the number of parent points making our method storage efficient: 884K v.s 1.06M averaged across all scenes in dataset~\cite{mip-nerf360}. We hypothesise that a configuration with attention can pull information from nearby points, allowing the method to reduce the number of points to store and represent the scene efficiently. Additionally, we ablated various configurations by varying $\lambda$ in Eq.~\ref{eqn:fusion} and  the number of heads in attention to find the best configuration (Tab.~\ref{table:attn}). We can see a right balance between the input features and attention features is important for best performance.

\noindent \textbf{Adaptive Tree Manipulation (ATM).}  We remove  Adaptive Tree Manipulation (ATM) and  add or delete the trees based only on parent nodes statistics and observe a drop in PSNR (Tab.~\ref{tab:core}) also visible in rendered images Fig.~\ref{fig:atm}.
This is because there is no mechanism to promote important children to parents that might hinder in populating trees correctly and failing to represent complex scenes effectively. On the other hand, this might also lead to deletion of important children nodes when deleting a parent. Our proposed ATM method can effectively alleviate these issues.
    Additionally, in Fig.~\ref{fig:promoted_parents} we show the point clouds of three scenes. Green points represent parents promoted from children during the optimization. Yellow points show parents that stayed parents during entire optimization. 
    It is clearly seen that the majority of the parent nodes are formed by promoting child nodes. Further note that parts with relatively flat geometry exhibit more yellow, while sophisticated geometry with high frequency details contain more green. Hence, ATM brings a further benefit of being able to fit sophisticated geometry better.  

\noindent \textbf{Inputs of MLP for attribute prediction} Tab.~\ref{table:arch} shows the analysis for the inputs used to predict the attributes. 
 We conduct the experiments of without using the distance from points to the center of AABB (denoted as \emph{w/o Distance}) and without using the 3D position information (denoted as \emph{w/o Position}) to predict attributes. It can be seen that position is very crucial for training while distance further improves the performance.
Lastly, we analyze the degrees of the SH encoding on the view directions by performing degree from 1 to 3 (denoted as \emph{SH D1} to \emph{SH D3}). Degree of 3 gives the best performance as it has more capacity to model complicated light effects.

\noindent \textbf{Scene Contraction.}  We analyze the proposed contraction technique applied on the unbounded scene (Tab.~\ref{table:arch} w/o Contract.).
Compared with \emph{Full}, 
we get inferior performance ($0.8$ PSNR drop), and tend to have training instability issues because the points occasionally move outside  the Axis-Aligned Bounding box. 

%% file: Table/comparison_table_full.tex
\begin{table*}[t] 
\small
  \caption{ Quantitative comparisons of our approach and other works evaluated on three widely used benchmark datasets, including Mip-NeRF 360$^\circ$ dataset~\cite{mip-nerf360}, Tanks\&Temples~\cite{tank}, and Deep Blending~\cite{db}.
  We report the image quality metrics, such as PSNR, SSIM, and LPIPS, and the required storage. We also report the on-device capability of each Gaussian Splatting based work (On-Device in the table), where {OOM denotes Out-of-Memory error and \checkmark denotes the real-time capability (>$30$ fps) on our tested device, \ie, iPhone14.}
  The evaluation results on other works are obtained from their papers.
  Compared with the methods that are capable to run on mobile devices, our models (Ours-C1, C2, C3) can obtain smaller model size with higher rendering quality (\ie, PSNR).
  }
  \label{tab:rendering_metrics}
  \centering
  \resizebox{1\linewidth}{!}{
  \begin{tabular}{llllllllllllll}
    \toprule
    Method  &   On-Device & \multicolumn{4}{c}{Mip-NeRF $360^\circ$ Dataset}   & \multicolumn{4}{c}{Tank\&Temples}  & \multicolumn{4}{c}{Deep Blending} \\
    \cmidrule(r){3-6}   \cmidrule(r){7-10} \cmidrule(r){11-14} 
      & & PSNR $\uparrow$ & SSIM $\uparrow$ & LPIPS  $\downarrow$  & Storage $\downarrow$  & PSNR $\uparrow$ & SSIM $\uparrow$ & LPIPS  $\downarrow$  & Storage $\downarrow$   & PSNR $\uparrow$ & SSIM $\uparrow$ & LPIPS  $\downarrow$  & Storage $\downarrow$   \\
    \midrule
    Mip-NeRF 360$^\circ$ & - & {29.23} & 0.844 & 0.207 & 8.6MB & 22.22 & 0.759 & 0.257 & 8.6MB & 29.40 & 0.901 & 0.245& 8.6MB  \\
    iNGP & - & 26.43 & 0.725 &  0.339  & 48MB & 21.72 & 0.723 & 0.330  & 48MB & 23.62 & 0.797 & 0.423 & 48MB  \\
    Plenoxels & - & 23.62 & 0.670 & 0.443 &  2100MB & 21.08 & 0.719 & 0.379  & 2300MB & 23.06 & 0.795 & 0.510 & 2700MB \\ \hline
    ScaffoldGS& OOM & {29.02} & 0.848 & 0.220  & 156MB &  {23.96} & {0.853} & {0.177}  & 87MB  &  {30.21} &  {0.906} & {0.254} &{66MB} \\
    3DGS & OOM &  {28.69} & {0.870} & {0.182} & 693MB & 23.14 & {0.841} & {0.183}  & 411MB & 29.41 &  {0.903} &  {0.243}  & 676MB\\
     \midrule
    LightGS & \checkmark & 28.45 &  {0.857} & {0.210} & {42.48MB} & 22.83 & 0.807 & 0.242  &  {22.43MB} &-&-&-&-\\
    CompactGS &\checkmark & {28.60} & {0.855} & {0.211}   & {46.98MB} & {23.32} & {0.831} & {0.201}    &{39.43MB} & {29.79} &{0.901} & {0.258} & {43.21MB}\\ \hline

     \textbf{Ours-C1} & \checkmark & {{28.45}} & {{0.837}} & {{0.235}} & {{23.40 MB}} & {23.19} & {0.810} &  {0.239}  &  {{22.00 MB}} &  {{29.32}} & {{0.895}} &  {0.282} &{22.90MB}  \\
     
      \textbf{Ours-C2} & \checkmark & {{28.86}} & {{0.851}} & {{0.217}} &  {{29.50 MB}} & {23.47} & {0.820} &  {0.228}  & {{29.05MB}} &  {{29.61}} & {{0.896}} &  {0.277} & {29.15MB}  \\
      
      \textbf{Ours-C3} & \checkmark & {{29.11}} & {{0.857}} & {{0.210}} &  {{35.60MB}} &    {23.82} & {0.829} &   {0.210}  &  {{35.32MB}} &  {{29.89}} & {{0.902}} &  {0.267} & {35.40MB}  \\

    \bottomrule
  \end{tabular}}

\label{tab:exp}
  
\end{table*}

%% file: figures/size_plot.tex
\begin{figure}[tbh]
\centering
\includegraphics[width=1\linewidth]{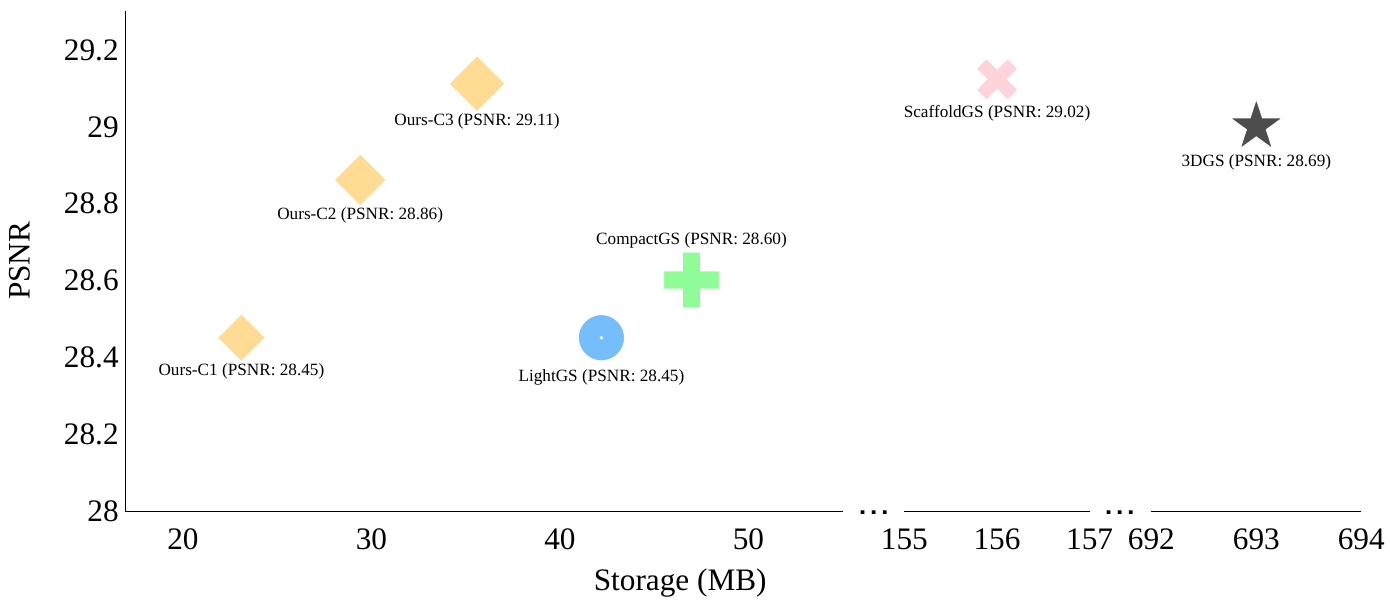}
\caption{We plot the PSNR score of several configurations of our method and prior works computed over the dataset introduced by~\cite{mip-nerf360}.}

\label{fig:size-plot}
\end{figure}

%% file: figures/exp.tex
\begin{figure*}
\centering
\includegraphics[width=1\textwidth]{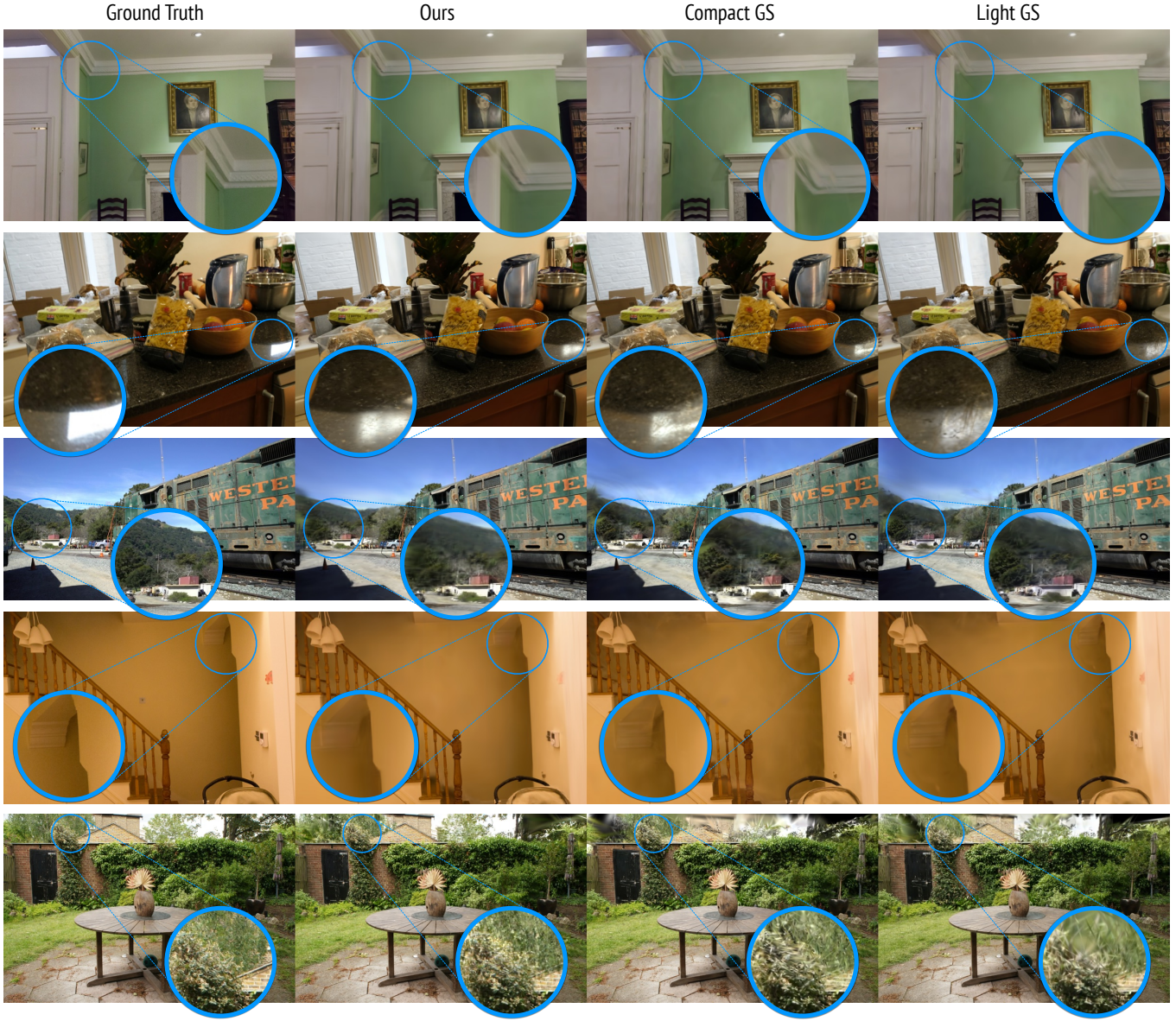}
\caption{Visual comparisons with methods offering efficient GS representations (\cite{compactgs, lightgs}). We magnified regions to show qualitative differences.
Our approach (C3) can render images with high-quality while greatly saving the storage. Zoom-in for greater detail.
}
\label{fig:eval}
\end{figure*}

%% file: figures/atm.tex
\begin{figure*}[t]
\includegraphics[width=\linewidth]{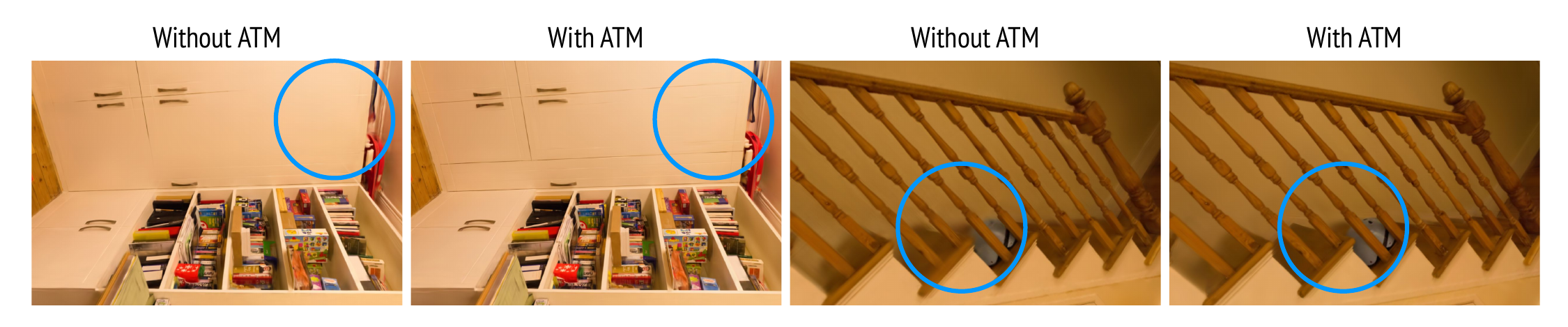}
\caption{Visual comparison of model trained with and without ATM. We can see that model trained without ATM fails to model intricate details in the scene.
}
\label{fig:atm}
\end{figure*}

%% file: Table/core_and_attn.tex
\begin{table}[ht]
\begin{minipage}{\linewidth}
  \centering
  \caption{Analysis of the key components in our framework. \emph{Full} denotes the full pipeline. We report the PSNR results for different settings.}\label{tab:core}
  \resizebox{1\linewidth}{!}{
\begin{tabular}{lccccc}
\toprule
&{FE}& {w/o Attn.}& {w/o ATM.}& {w/o Contract.}   &\textbf{Full} 
\\
\midrule
 \texttt{Bicycle} & 19.72 & 22.80 & 22.66& 22.88 & 23.68 \\
 \midrule
\texttt{Playroom} & 23.36 & 28.74 & 28.65 & -   & 29.27 \\
\bottomrule
 
\end{tabular}}
\end{minipage}%
\hfill

\label{tab:ablations-components}
\end{table}

%% file: Table/attn.tex
\begin{table}[ht]
\centering
\small
\caption{Analysis of self-attention. We show the PSNR for the attention with different number of heads (H) and attention head dimension (F), and different $\lambda$ in Eqn.~\ref{eqn:fusion}.}
\label{table:attn}

\begin{tabular}{lccccccc}
\toprule

&{H1-F32}
& {H2-F16} 
& {H4-F8} 
& {$\lambda$=0.1}
& {$\lambda$=0.5}
& {$\lambda$=1.0}

\\
\midrule
 \texttt{Bicycle} & 23.68 & 23.31 & 23.20 & 23.55 & 23.68 & 23.53 \\

 \midrule
\texttt{Playroom} & 29.27 & 29.15 & 29.10 & 29.14 & 29.27 & 29.22 \\
\bottomrule
 
\end{tabular}

\end{table}

%% file: Table/arch.tex
\begin{table}[ht]
\centering
\small
\caption{Analysis of the inputs used for attributes prediction.}
\label{table:arch}

\begin{tabular}{lcccccccc}
\toprule
 
& {w/o Distance}
& {w/o Position}
& {SH D1}
& {SH D2}
& {SH D3}

\\
\midrule
 \texttt{Bicycle} &23.32 & 9.72 & 23.05 & 23.60 & 23.68 \\

 \midrule
\texttt{Playroom}  & 29.18 & 6.19 & 29.23 & 29.15 & 29.27 \\

\bottomrule 
\end{tabular}

\end{table}

%% file: figures/promoted_parents.tex
\begin{figure}[t]
\includegraphics[width=\linewidth]{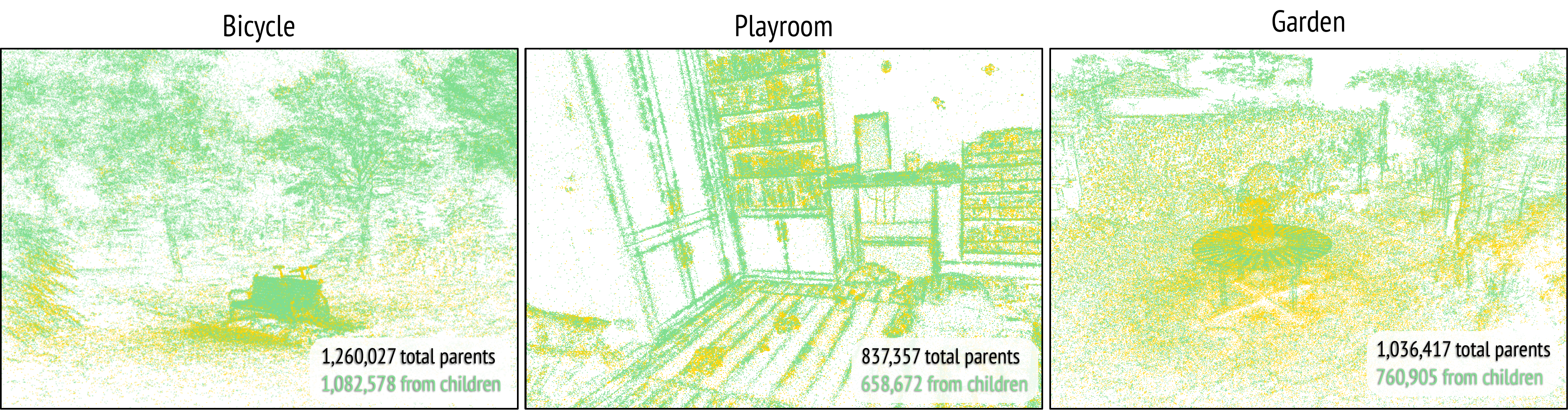}
\caption{The effect of Adaptive Tree Manipulation (ATM). Yellow points indicate the splats who have not changed the parent status during entire optimization. Green points represent former children that have been promoted to parents. Around 80\% of parents are from our ATM operation. 
}
\label{fig:promoted_parents}
\end{figure}

%% file: sec/5_conclusion.tex
\section{Conclusion}
This paper introduces predictive 3D Gaussian splats, a lightweight representation that dramatically reduces storage for large-scale scenes compared to 3DGS, while maintaining high-fidelity rendering results. To build the framework, we introduce several new techniques. For example, we propose an efficient \emph{parent}-\emph{children} structure that only requires saving \emph{parent} points. The \emph{children} points and most Gaussian attributes can be estimated during rendering by utilizing \emph{parent} points. Additionally, we leverage a hash grid and self-attention on aggregated features to enforce connectivity for \emph{parent} and \emph{children} nodes. We conduct extensive experiments on benchmark datasets to validate our design and demonstrate the our advantages of storage saving and high-quality novel view synthesis.

%% file: sec/supp.tex
\section{Implementation Details}

In this section, we provide more details for our training. We first provide the learning rates used during the traing in \ref{lr}. Next we show the visual illustration and the implementation details of the contraction in Sec.~\ref{sec:contract}. Then, we analyze the effectiveness of the warm-up training strategy employed in our method in Sec.~\ref{sec:warm}. Lastly, we discuss the pre-filtering of \emph{parent} points and its implementation in Sec.~\ref{sec:cull}.

\subsection{Settings of Learning Rate} \label{lr}
We employ different learning schedules for different modules. For the hash grid, we start with a learning rate of $2e^{-3}$ and end with a rate of $2e^{-5}$. For opacity, we start with $1e^{-3}$ and end with $2e^{-5}$. The scale and rotation parameters utilize a constant learning rate of $1e^{-4}$. Additionally, we maintain a constant learning rate of $2e^{-4}$ for the attention module. We apply a standard exponential decay scheduling \cite{3dgs, plenoxels} to all modules. 

\subsection{Details for Contraction} \label{sec:contract}
\input{figures/aabb}
We illustrate the process of contraction (described in Sec. 3.5 of the main paper) in Fig.~\ref{fig:aabb} and Alg.~\ref{alg:cap}.  We calculate the inscribed and circumscribed spheres (\ie, $S_{inner}$ and $S_{outer}$) with radius $R_{inner}$ and $R_{outer}$  of the initialized Axis-Aligned Bounding Box (AABB), which is estimated from the point cloud generated from COLMAP~\cite{colmap}. The estimated AABB is the circumscribed cube of the outer sphere $S_{outer}$. Points falling outside of the outer sphere are brought back to $S_{outer}$.

\begin{algorithm}
\caption{AABB Estimation and Contraction}\label{alg:cap}
\begin{algorithmic}
\Require Initialized AABB: $AABB_{init}$, point cloud: $PC$
\State $S_{inner} \gets \text{Inscribed sphere of }AABB_{init}$    \Comment{centered at $O$}
\State $R_{inner} \gets \text{Radius of }S_{inner}$
\State $S_{outer} \gets \text{Circumscribed sphere of }AABB_{init}$   \Comment{centered at $O$}
\State $R_{outer} \gets \text{Radius of }S_{outer}$
\State $AABB_{est} \gets \text{Circumscribed cube of }S_{outer}$
\\   
\For {$p$ in $PC$}               \Comment{contract the points}        
\If{ $\|{p - O}\| \leq R_{inner}$}
    \State $p \gets p$
\ElsIf{ $\|{p - O}\| > R_{inner}$}
    \State $ p \gets \Bigl(R_{outer} - \frac{1}{\|{p - O}\|}\Bigr) \Bigl(\frac{p - O}{\|{p - O}\|} \Bigr) + O $  \Comment{infinity is on $S_{outer}$}   
\EndIf
\EndFor
\end{algorithmic}
\end{algorithm}

\input{figures/warmup}
\subsection{Analysis of Warm-up} \label{sec:warm}
We run two experiments on the \texttt{Garden}~\cite{mip-nerf360} scene in 10K steps to show the effectiveness of the warm-up in our method.
We have found that using warm-up in training with low resolution images at early stages helps the points populate the empty areas, especially when the COLMAP~\cite{colmap} initialization is poor.
Fig.~\ref{fig:warmup} shows the point cloud and corresponding rendered images from different training approaches. As can be seen,
the warm-up training (second row) has a better reconstruction and rendering quality for the background scene, which is poorly initialized from COLMAP~\cite{colmap}.

\subsection{View Frustum Culling}\label{sec:cull}
We apply pre-filtering on the \emph{parent} points before querying features for attributes prediction by culling the view frustum with depth, leading to the computation reduction and the training speedup.
We empirically observe that $15\%$ - $25\%$ points are removed across scenes. The implementation is shown in Alg.~\ref{alg:culiing}.

\begin{algorithm}
\caption{View Frustum Culling}\label{alg:culiing}
\begin{algorithmic}
\Require  points $P: N \times 3$, view matrix $M: 4\times 4$
\State $P_{homo} \gets Concat(P, ones) $ \Comment{$P_{homo}: N \times 4$}
\State $P_{view} \gets M * (P_{homo})^ T$ \Comment{$*$ is matrix multiplication}
\State $mask \gets P_{view}[2, :] > 0.201 $ \Comment{depth $> 0.201$}
\State $P_{filtered} \gets P[mask]$
\end{algorithmic}
\end{algorithm}

\section{Inference Phase Optimization}

Predicting the attributes brings overheads during the inference. However, we note that only the attribute color is view-dependent and the rest remain the same for all frames. Therefore, to minimize the computation cost, we opt to run the color prediction (small MLPs) only and the rest of the attributes can be retrieved from the first frame. This technique enables us method the real-time rendering on the mobile phones.

\section{Per-scene Quantitative Results}

We provide the per-scene results on the benchmark datasets. Tab.~\ref{tab:360_psnr} shows the results on the Mip-NeRF $360^\circ$ dataset~\cite{mip-nerf360}. Tab.~\ref{tab:tt and db} demonstrates the results on the Tank\&Temples dataset~\cite{tank}  and the Deep Blending dataset~\cite{db}.
We report the per-scene storage (in MB), the number of \emph{parents} and \emph{children}, and the metrics for image quality evaluation.
\input{Table/perscene-metric}

\clearpage

%% file: figures/aabb.tex
\begin{figure}[t]
\includegraphics[width=1\linewidth]{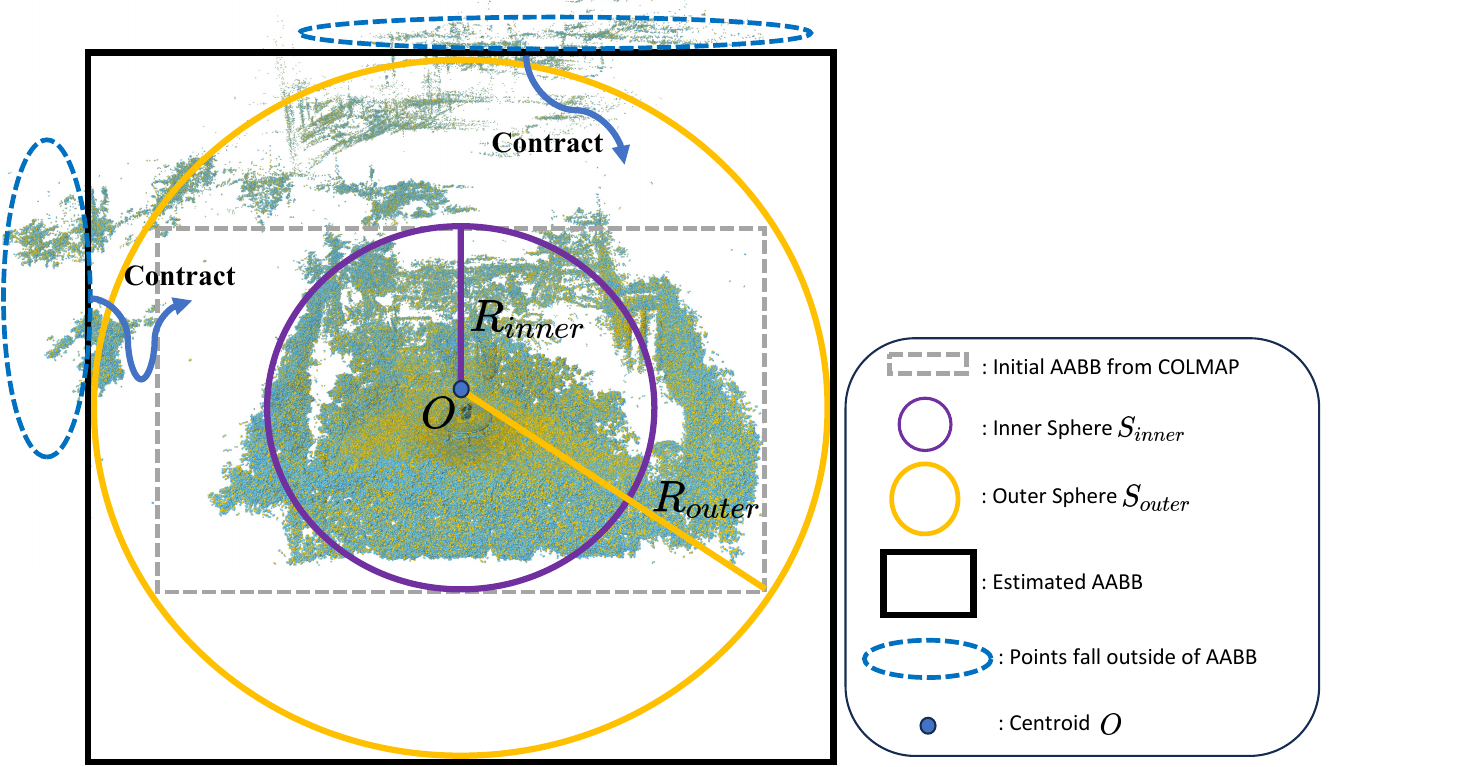}
\caption{{Illustration for our implemented contraction.}
}
\label{fig:aabb}
\end{figure}

%% file: figures/warmup.tex
\begin{figure}[t]
\centering
\includegraphics[width=\linewidth]{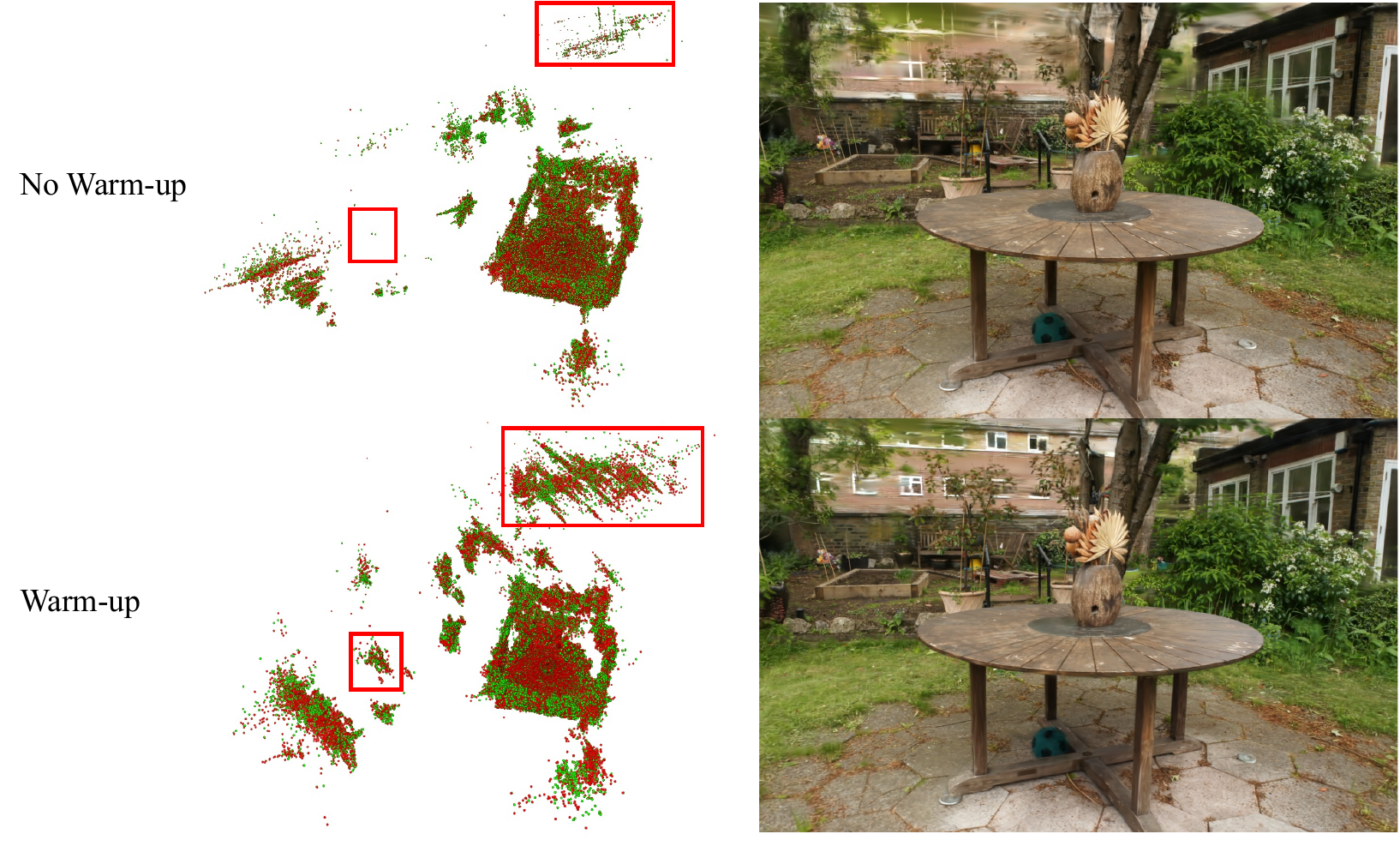}
\caption{Analysis of warm-up. \emph{First row}:training without warm-up. \emph{Second row}: training with warm-up. \emph{Left column}: points distribution. \emph{Right column}: rendered images with the point cloud. }
\label{fig:warmup}
\end{figure}

%% file: Table/perscene-metric.tex
\begin{table}[h]
  \caption{Per-scene metrics for our approach on the Mip-NeRF $360^\circ$ dataset~\cite{mip-nerf360}.}
  \label{tab:360_psnr}
  \centering
  \resizebox{1\linewidth}{!}{
  \begin{tabular}{lllllllll}
    \toprule
 
    \multirow{8}{6em}{Compact GS} & Metric & Garden & Bicycle & Stump &  Room & Counter & Kitchen & Bonsai \\
    \midrule

     & PSNR &  26.81 & 24.77 & 26.46 & 30.88 & 28.71 & 30.480 & 32.08  \\
     & SSIM & 0.832 & 0.723 & 0.757 & 0.919 & 0.902 & 0.919 & 0.939\\
     & LPIPS  & 0.161 & 0.286 & 0.278 & 0.209 & 0.205 & 0.131 & 0.193 \\

     & Storage (MB) & 62.78 & 62.99 & 54.66 & 34.21 & 34.34 & 44.45 & 35.44  \\
    \midrule

     \multirow{5}{6em}{Light GS} 
     & PSNR &  26.73 & 24.96 & 26.70 & 31.27 & 28.11 & 30.40 & 31.01  \\
     & SSIM& 0.836 & 0.738 & 0.768 & 0.926 & 0.893 & 0.914 & 0.944\\
     & LPIPS & 0.155 & 0.265 & 0.261 & 0.220 & 0.218 & 0.147 & 0.204 \\
     & Storage (MB) & - & - & - & - & - & - & -  \\
    \midrule

     \multirow{5}{6em}{Scaffold GS} 

     & PSNR &  27.17 & 24.50 & 26.27 & 31.93 & 29.34 & 31.30 & 32.70 \\
     & SSIM & 0.842 & 0.705 & 0.784 & 0.925 & 0.914 & 0.928 & 0.946\\
     & LPIPS & 0.146 & 0.306 & 0.284 & 0.202 & 0.191 & 0.126 & 0.185\\
     & Storage (MB) & 271.00 & 248.00 & 493.00 & 133.00 & 194.00 & 173.00 & 258.00\\
    \midrule

     \multirow{5}{6em}{3D GS} 

     & PSNR &  27.25 & 25.10 & 26.66 & 31.50 & 29.11 & 31.53 & 32.16 \\
     & SSIM & 0.856 & 0.747 & 0.756  & 0.925 & 0.914 & 0.932 & 0.946\\
     & LPIPS & 0.122 & 0.244 & 0.243 & 0.198 & 0.184 & 0.117 & 0.181\\
     & Storage (MB) & 1331.33 & 1350.78 & 1073.60 & 350.14 & 276.52 & 411.76 & 295.08\\
    \midrule
    \midrule
     \multirow{5}{6em}{Ours-C1} 

     & PSNR &  27.17 & 24.32 & 25.75 & 31.62 & 28.54 & 30.47 & 31.32\\
     & SSIM & 0.832 & 0.672 & 0.768 & 0.913 & 0.889 & 0.910 & 0.923\\
     & LPIPS & 0.169 & 0.355 & 0.312 & 0.229 & 0.226 & 0.147 & 0.206\\
     & Storage (MB) & 26.90 & 25.71 & 34.84 & 16.85 & 17.92 & 23.68 & 17.74\\
    \midrule
    
    \multirow{5}{6em}{Ours-C2} 

     & PSNR &  27.38 & 24.78 & 26.41 & 31.82 & 28.75 & 30.71 & 32.14 \\
     & SSIM & 0.842 & 0.701 & 0.751 & 0.916 & 0.894 & 0.913 & 0.935 \\
     & LPIPS  & 0.156 & 0.325 & 0.260 & 0.224 & 0.218 & 0.146 & 0.192\\
     
     & Storage (MB) & 33.01 & 31.72 & 41.09 & 23.02 & 24.04 & 30.05 & 23.61 \\
    \midrule
    
     \multirow{5}{6em}{Ours-C3} 

     & PSNR &  27.63 & 24.90 & 26.43 & 31.84 & 29.10 & 31.27 & 32.67  \\
     & SSIM & 0.847 & 0.717 & 0.753 & 0.917 & 0.900 & 0.918 & 0.941 \\
     & LPIPS  & 0.147 & 0.303 & 0.267 & 0.220 & 0.212 & 0.137 & 0.186 \\
     
     & Storage (MB) & 39.40 & 37.81 & 47.24 & 28.95 & 30.02 & 35.92 & 29.84 \\

     \midrule
     & \# of Parents & 1.20M & 1.06M & 1.86M & 330K & 419K & 913K & 403K  \\
     & \# of Children ($k$) & 2 & 2 & 2 & 2 & 2 &  1 & 2 \\
    \bottomrule
  \end{tabular}}
\end{table}

\begin{table}[!h]
 \small
  \caption{Per-scene metrics for our approach on the Tank\&Temples dataset~\cite{tank} and the Deep Blending dataset~\cite{db}.}
  \label{tab:tt and db}
  \centering
  \resizebox{0.8\linewidth}{!}{
  \begin{tabular}{llllll}
    \toprule
      \multirow{10}{6em}{Compact GS} & Metric & \multicolumn{2}{c}{Tank\&Temples}  & \multicolumn{2}{c}{Deep Blending} \\
    \cmidrule(r){3-4}   \cmidrule(r){5-6} 
       & & Truck  & Train  & Drjohnson  & Playroom \\
    \midrule
    & PSNR & 25.070 & 21.560  & 29.260 & 30.320 \\
    & SSIM &0.871 & 0.792  & 0.9000 & 0.902\\
    &LPIPS & 0.163 & 0.240  & 0.258 & 0.258 \\
    & Storage (MB) & 41.57 & 37.29  & 47.98 & 38.45\\
     \midrule

      \multirow{4}{6em}{Light GS} 

    & PSNR & 24.561 & 21.095 & - & -  \\
    &SSIM & 0.855 & 0.760   & - & -  \\
    &LPIPS & 0.188 & 0.296  & - & -\\
    &Storage (MB) & - & -  & - & -   \\
     \midrule
      \multirow{4}{6em}{Scaffold GS} 

   & PSNR & 25.77 & 22.15 & 29.80 & 30.62\\
    &SSIM& 0.883 & 0.822  & 0.907 & 0.904\\
    &LPIPS  & 0.147 & 0.206  & 0.250 & 0.258\\
        &Storage (MB) & 107.00 & 66.00 & 69.00 & 63.00 \\
     \midrule
      \multirow{4}{6em}{3D GS} 

    & PSNR & 25.350 & 22.070  & 29.060 & 29.870\\
    &SSIM  & 0.878 & 0.812  & 0.899 & 0.901\\
    &LPIPS  & 0.148 & 0.208 & 0.247 & 0.247\\
    &Storage (MB) & 608.70 & 255.82 & 773.61 & 553.03 \\
     \midrule
     \midrule
     
      \multirow{5}{6em}{Ours-C1} 

     & PSNR & 24.93 & 21.44  & 28.89 & 29.75 \\
    &SSIM  & 0.856 & 0.763  & 0.894 & 0.895\\
    &LPIPS  & 0.196 & 0.283 & 0.280 & 0.284 \\
        &Storage (MB) & 23.11 & 20.90  & 23.59  & 22.21  \\
     \midrule
      \multirow{5}{6em}{Ours-C2} 

    & PSNR & 25.22 & 21.72  & 28.93 & 30.28 \\
    &SSIM & 0.862 & 0.777  & 0.902 & 0.902\\
    &LPIPS & 0.184 & 0.272 & 0.287 & 0.268 \\
        &Storage (MB) & 30.73 & 27.36 & 29.84 & 28.46\\
     \midrule
      \multirow{5}{6em}{Ours-C3} 

      & PSNR & 25.45 & 22.18 & 29.34 & 30.44\\
    &SSIM  & 0.866 & 0.792 & 0.898 & 0.905  \\
    &LPIPS  & 0.182 & 0.240 & 0.270 & 0.265 \\
        &Storage (MB) &  36.01 & 34.63  & 35.80  & 35.00  \\
     \midrule
     &\# of Parents & 1M & 900K  & 900K  & 834K \\
    &\# of Children ($k$) & 1 & 1 & 2 & 2 \\

    \bottomrule
  \end{tabular}}
\end{table}

%% file: main.bbl

\begin{thebibliography}{58}


\ifx \showCODEN    \undefined \def \showCODEN     #1{\unskip}     \fi
\ifx \showDOI      \undefined \def \showDOI       #1{#1}\fi
\ifx \showISBNx    \undefined \def \showISBNx     #1{\unskip}     \fi
\ifx \showISBNxiii \undefined \def \showISBNxiii  #1{\unskip}     \fi
\ifx \showISSN     \undefined \def \showISSN      #1{\unskip}     \fi
\ifx \showLCCN     \undefined \def \showLCCN      #1{\unskip}     \fi
\ifx \shownote     \undefined \def \shownote      #1{#1}          \fi
\ifx \showarticletitle \undefined \def \showarticletitle #1{#1}   \fi
\ifx \showURL      \undefined \def \showURL       {\relax}        \fi
\providecommand\bibfield[2]{#2}
\providecommand\bibinfo[2]{#2}
\providecommand\natexlab[1]{#1}
\providecommand\showeprint[2][]{arXiv:#2}

\bibitem[bot(2005)]%
        {botsch2005high}
 \bibinfo{year}{2005}\natexlab{}.
\newblock \showarticletitle{High-quality surface splatting on today's GPUs}. In \bibinfo{booktitle}{\emph{Proceedings Eurographics/IEEE VGTC Symposium Point-Based Graphics, 2005.}} IEEE, \bibinfo{pages}{17--141}.
\newblock


\bibitem[Aliev et~al\mbox{.}(2020)]%
        {aliev2020neural}
\bibfield{author}{\bibinfo{person}{Kara-Ali Aliev}, \bibinfo{person}{Artem Sevastopolsky}, \bibinfo{person}{Maria Kolos}, \bibinfo{person}{Dmitry Ulyanov}, {and} \bibinfo{person}{Victor Lempitsky}.} \bibinfo{year}{2020}\natexlab{}.
\newblock \showarticletitle{Neural point-based graphics}. In \bibinfo{booktitle}{\emph{Computer Vision--ECCV 2020: 16th European Conference, Glasgow, UK, August 23--28, 2020, Proceedings, Part XXII 16}}. Springer, \bibinfo{pages}{696--712}.
\newblock


\bibitem[Barron et~al\mbox{.}(2022a)]%
        {mip-nerf360}
\bibfield{author}{\bibinfo{person}{Jonathan~T Barron}, \bibinfo{person}{Ben Mildenhall}, \bibinfo{person}{Dor Verbin}, \bibinfo{person}{Pratul~P Srinivasan}, {and} \bibinfo{person}{Peter Hedman}.} \bibinfo{year}{2022}\natexlab{a}.
\newblock \showarticletitle{Mip-nerf 360: Unbounded anti-aliased neural radiance fields}. In \bibinfo{booktitle}{\emph{Proceedings of the IEEE/CVF Conference on Computer Vision and Pattern Recognition}}. \bibinfo{pages}{5470--5479}.
\newblock


\bibitem[Barron et~al\mbox{.}(2022b)]%
        {mipnerf}
\bibfield{author}{\bibinfo{person}{Jonathan~T. Barron}, \bibinfo{person}{Ben Mildenhall}, \bibinfo{person}{Dor Verbin}, \bibinfo{person}{Pratul~P. Srinivasan}, {and} \bibinfo{person}{Peter Hedman}.} \bibinfo{year}{2022}\natexlab{b}.
\newblock \showarticletitle{Mip-NeRF 360: Unbounded Anti-Aliased Neural Radiance Fields}.
\newblock \bibinfo{journal}{\emph{CVPR}} (\bibinfo{year}{2022}).
\newblock


\bibitem[Buehler et~al\mbox{.}(2023)]%
        {buehler2023unstructured}
\bibfield{author}{\bibinfo{person}{Chris Buehler}, \bibinfo{person}{Michael Bosse}, \bibinfo{person}{Leonard McMillan}, \bibinfo{person}{Steven Gortler}, {and} \bibinfo{person}{Michael Cohen}.} \bibinfo{year}{2023}\natexlab{}.
\newblock \showarticletitle{Unstructured lumigraph rendering}.
\newblock In \bibinfo{booktitle}{\emph{Seminal Graphics Papers: Pushing the Boundaries, Volume 2}}. \bibinfo{pages}{497--504}.
\newblock


\bibitem[Cao et~al\mbox{.}(2023)]%
        {mobiler2l}
\bibfield{author}{\bibinfo{person}{Junli Cao}, \bibinfo{person}{Huan Wang}, \bibinfo{person}{Pavlo Chemerys}, \bibinfo{person}{Vladislav Shakhrai}, \bibinfo{person}{Ju Hu}, \bibinfo{person}{Yun Fu}, \bibinfo{person}{Denys Makoviichuk}, \bibinfo{person}{Sergey Tulyakov}, {and} \bibinfo{person}{Jian Ren}.} \bibinfo{year}{2023}\natexlab{}.
\newblock \showarticletitle{Real-Time Neural Light Field on Mobile Devices}. In \bibinfo{booktitle}{\emph{Proceedings of the IEEE/CVF Conference on Computer Vision and Pattern Recognition}}. \bibinfo{pages}{8328--8337}.
\newblock


\bibitem[Chen et~al\mbox{.}(2023b)]%
        {chen2023dictionary}
\bibfield{author}{\bibinfo{person}{Anpei Chen}, \bibinfo{person}{Zexiang Xu}, \bibinfo{person}{Xinyue Wei}, \bibinfo{person}{Siyu Tang}, \bibinfo{person}{Hao Su}, {and} \bibinfo{person}{Andreas Geiger}.} \bibinfo{year}{2023}\natexlab{b}.
\newblock \showarticletitle{Dictionary fields: Learning a neural basis decomposition}.
\newblock \bibinfo{journal}{\emph{ACM Transactions on Graphics (TOG)}} \bibinfo{volume}{42}, \bibinfo{number}{4} (\bibinfo{year}{2023}), \bibinfo{pages}{1--12}.
\newblock


\bibitem[Chen and Williams(2023)]%
        {chen2023view}
\bibfield{author}{\bibinfo{person}{Shenchang~Eric Chen} {and} \bibinfo{person}{Lance Williams}.} \bibinfo{year}{2023}\natexlab{}.
\newblock \showarticletitle{View interpolation for image synthesis}.
\newblock In \bibinfo{booktitle}{\emph{Seminal Graphics Papers: Pushing the Boundaries, Volume 2}}. \bibinfo{pages}{423--432}.
\newblock


\bibitem[Chen et~al\mbox{.}(2023a)]%
        {mobilenerf}
\bibfield{author}{\bibinfo{person}{Zhiqin Chen}, \bibinfo{person}{Thomas Funkhouser}, \bibinfo{person}{Peter Hedman}, {and} \bibinfo{person}{Andrea Tagliasacchi}.} \bibinfo{year}{2023}\natexlab{a}.
\newblock \showarticletitle{Mobilenerf: Exploiting the polygon rasterization pipeline for efficient neural field rendering on mobile architectures}. In \bibinfo{booktitle}{\emph{Proceedings of the IEEE/CVF Conference on Computer Vision and Pattern Recognition}}. \bibinfo{pages}{16569--16578}.
\newblock


\bibitem[Choi et~al\mbox{.}(2019)]%
        {choi2019extreme}
\bibfield{author}{\bibinfo{person}{Inchang Choi}, \bibinfo{person}{Orazio Gallo}, \bibinfo{person}{Alejandro Troccoli}, \bibinfo{person}{Min~H Kim}, {and} \bibinfo{person}{Jan Kautz}.} \bibinfo{year}{2019}\natexlab{}.
\newblock \showarticletitle{Extreme view synthesis}. In \bibinfo{booktitle}{\emph{Proceedings of the IEEE/CVF International Conference on Computer Vision}}. \bibinfo{pages}{7781--7790}.
\newblock


\bibitem[Fan et~al\mbox{.}(2024)]%
        {lightgs}
\bibfield{author}{\bibinfo{person}{Zhiwen Fan}, \bibinfo{person}{Kevin Wang}, \bibinfo{person}{Kairun Wen}, \bibinfo{person}{Zehao Zhu}, \bibinfo{person}{Dejia Xu}, {and} \bibinfo{person}{Zhangyang Wang}.} \bibinfo{year}{2024}\natexlab{}.
\newblock \bibinfo{title}{LightGaussian: Unbounded 3D Gaussian Compression with 15x Reduction and 200+ FPS}.
\newblock
\newblock
\showeprint[arxiv]{2311.17245}~[cs.CV]


\bibitem[Feng et~al\mbox{.}(2022)]%
        {feng2022neural}
\bibfield{author}{\bibinfo{person}{Wanquan Feng}, \bibinfo{person}{Jin Li}, \bibinfo{person}{Hongrui Cai}, \bibinfo{person}{Xiaonan Luo}, {and} \bibinfo{person}{Juyong Zhang}.} \bibinfo{year}{2022}\natexlab{}.
\newblock \showarticletitle{Neural points: Point cloud representation with neural fields for arbitrary upsampling}. In \bibinfo{booktitle}{\emph{Proceedings of the IEEE/CVF Conference on Computer Vision and Pattern Recognition}}. \bibinfo{pages}{18633--18642}.
\newblock


\bibitem[Flynn et~al\mbox{.}(2019)]%
        {flynn2019deepview}
\bibfield{author}{\bibinfo{person}{John Flynn}, \bibinfo{person}{Michael Broxton}, \bibinfo{person}{Paul Debevec}, \bibinfo{person}{Matthew DuVall}, \bibinfo{person}{Graham Fyffe}, \bibinfo{person}{Ryan Overbeck}, \bibinfo{person}{Noah Snavely}, {and} \bibinfo{person}{Richard Tucker}.} \bibinfo{year}{2019}\natexlab{}.
\newblock \showarticletitle{Deepview: View synthesis with learned gradient descent}. In \bibinfo{booktitle}{\emph{Proceedings of the IEEE/CVF Conference on Computer Vision and Pattern Recognition}}. \bibinfo{pages}{2367--2376}.
\newblock


\bibitem[Fridovich-Keil et~al\mbox{.}(2022)]%
        {fridovich2022plenoxels}
\bibfield{author}{\bibinfo{person}{Sara Fridovich-Keil}, \bibinfo{person}{Alex Yu}, \bibinfo{person}{Matthew Tancik}, \bibinfo{person}{Qinhong Chen}, \bibinfo{person}{Benjamin Recht}, {and} \bibinfo{person}{Angjoo Kanazawa}.} \bibinfo{year}{2022}\natexlab{}.
\newblock \showarticletitle{Plenoxels: Radiance fields without neural networks}. In \bibinfo{booktitle}{\emph{Proceedings of the IEEE/CVF Conference on Computer Vision and Pattern Recognition}}. \bibinfo{pages}{5501--5510}.
\newblock


\bibitem[Garbin et~al\mbox{.}(2021)]%
        {garbin2021fastnerf}
\bibfield{author}{\bibinfo{person}{Stephan~J Garbin}, \bibinfo{person}{Marek Kowalski}, \bibinfo{person}{Matthew Johnson}, \bibinfo{person}{Jamie Shotton}, {and} \bibinfo{person}{Julien Valentin}.} \bibinfo{year}{2021}\natexlab{}.
\newblock \showarticletitle{Fastnerf: High-fidelity neural rendering at 200fps}. In \bibinfo{booktitle}{\emph{Proceedings of the IEEE/CVF International Conference on Computer Vision}}. \bibinfo{pages}{14346--14355}.
\newblock


\bibitem[Greene(1986)]%
        {greene1986environment}
\bibfield{author}{\bibinfo{person}{Ned Greene}.} \bibinfo{year}{1986}\natexlab{}.
\newblock \showarticletitle{Environment mapping and other applications of world projections}.
\newblock \bibinfo{journal}{\emph{IEEE computer graphics and Applications}} \bibinfo{volume}{6}, \bibinfo{number}{11} (\bibinfo{year}{1986}), \bibinfo{pages}{21--29}.
\newblock


\bibitem[Gross and Pfister(2011)]%
        {gross2011point}
\bibfield{author}{\bibinfo{person}{Markus Gross} {and} \bibinfo{person}{Hanspeter Pfister}.} \bibinfo{year}{2011}\natexlab{}.
\newblock \bibinfo{booktitle}{\emph{Point-based graphics}}.
\newblock \bibinfo{publisher}{Elsevier}.
\newblock


\bibitem[Gupta et~al\mbox{.}(2024)]%
        {lightspeed}
\bibfield{author}{\bibinfo{person}{Aarush Gupta}, \bibinfo{person}{Junli Cao}, \bibinfo{person}{Chaoyang Wang}, \bibinfo{person}{Ju Hu}, \bibinfo{person}{Sergey Tulyakov}, \bibinfo{person}{Jian Ren}, {and} \bibinfo{person}{L{\'a}szl{\'o} Jeni}.} \bibinfo{year}{2024}\natexlab{}.
\newblock \showarticletitle{LightSpeed: Light and Fast Neural Light Fields on Mobile Devices}.
\newblock \bibinfo{journal}{\emph{Advances in Neural Information Processing Systems}}  \bibinfo{volume}{36} (\bibinfo{year}{2024}).
\newblock


\bibitem[Hedman et~al\mbox{.}(2018)]%
        {db}
\bibfield{author}{\bibinfo{person}{Peter Hedman}, \bibinfo{person}{Julien Philip}, \bibinfo{person}{True Price}, \bibinfo{person}{Jan-Michael Frahm}, \bibinfo{person}{George Drettakis}, {and} \bibinfo{person}{Gabriel Brostow}.} \bibinfo{year}{2018}\natexlab{}.
\newblock \showarticletitle{Deep Blending for Free-viewpoint Image-based Rendering}.
\newblock  \bibinfo{volume}{37}, \bibinfo{number}{6} (\bibinfo{year}{2018}), \bibinfo{pages}{257:1--257:15}.
\newblock


\bibitem[Insafutdinov and Dosovitskiy(2018)]%
        {insafutdinov2018unsupervised}
\bibfield{author}{\bibinfo{person}{Eldar Insafutdinov} {and} \bibinfo{person}{Alexey Dosovitskiy}.} \bibinfo{year}{2018}\natexlab{}.
\newblock \showarticletitle{Unsupervised learning of shape and pose with differentiable point clouds}.
\newblock \bibinfo{journal}{\emph{Advances in neural information processing systems}}  \bibinfo{volume}{31} (\bibinfo{year}{2018}).
\newblock


\bibitem[Kalantari et~al\mbox{.}(2016)]%
        {kalantari2016learning}
\bibfield{author}{\bibinfo{person}{Nima~Khademi Kalantari}, \bibinfo{person}{Ting-Chun Wang}, {and} \bibinfo{person}{Ravi Ramamoorthi}.} \bibinfo{year}{2016}\natexlab{}.
\newblock \showarticletitle{Learning-based view synthesis for light field cameras}.
\newblock \bibinfo{journal}{\emph{ACM Transactions on Graphics (TOG)}} \bibinfo{volume}{35}, \bibinfo{number}{6} (\bibinfo{year}{2016}), \bibinfo{pages}{1--10}.
\newblock


\bibitem[Kerbl et~al\mbox{.}(2023)]%
        {3dgs}
\bibfield{author}{\bibinfo{person}{Bernhard Kerbl}, \bibinfo{person}{Georgios Kopanas}, \bibinfo{person}{Thomas Leimk{\"u}hler}, {and} \bibinfo{person}{George Drettakis}.} \bibinfo{year}{2023}\natexlab{}.
\newblock \showarticletitle{3D Gaussian Splatting for Real-Time Radiance Field Rendering}.
\newblock \bibinfo{journal}{\emph{ACM Transactions on Graphics}} \bibinfo{volume}{42}, \bibinfo{number}{4} (\bibinfo{year}{2023}).
\newblock


\bibitem[Knapitsch et~al\mbox{.}(2017)]%
        {tank}
\bibfield{author}{\bibinfo{person}{Arno Knapitsch}, \bibinfo{person}{Jaesik Park}, \bibinfo{person}{Qian-Yi Zhou}, {and} \bibinfo{person}{Vladlen Koltun}.} \bibinfo{year}{2017}\natexlab{}.
\newblock \showarticletitle{Tanks and Temples: Benchmarking Large-Scale Scene Reconstruction}.
\newblock \bibinfo{journal}{\emph{ACM Transactions on Graphics}} \bibinfo{volume}{36}, \bibinfo{number}{4} (\bibinfo{year}{2017}).
\newblock


\bibitem[Kopanas et~al\mbox{.}(2021)]%
        {kopanas2021point}
\bibfield{author}{\bibinfo{person}{Georgios Kopanas}, \bibinfo{person}{Julien Philip}, \bibinfo{person}{Thomas Leimk{\"u}hler}, {and} \bibinfo{person}{George Drettakis}.} \bibinfo{year}{2021}\natexlab{}.
\newblock \showarticletitle{Point-Based Neural Rendering with Per-View Optimization}. In \bibinfo{booktitle}{\emph{Computer Graphics Forum}}, Vol.~\bibinfo{volume}{40}. Wiley Online Library, \bibinfo{pages}{29--43}.
\newblock


\bibitem[Lee et~al\mbox{.}(2024)]%
        {compactgs}
\bibfield{author}{\bibinfo{person}{Joo~Chan Lee}, \bibinfo{person}{Daniel Rho}, \bibinfo{person}{Xiangyu Sun}, \bibinfo{person}{Jong~Hwan Ko}, {and} \bibinfo{person}{Eunbyung Park}.} \bibinfo{year}{2024}\natexlab{}.
\newblock \bibinfo{title}{Compact 3D Gaussian Representation for Radiance Field}.
\newblock
\newblock
\showeprint[arxiv]{2311.13681}~[cs.CV]


\bibitem[Levoy and Hanrahan(2023)]%
        {levoy2023light}
\bibfield{author}{\bibinfo{person}{Marc Levoy} {and} \bibinfo{person}{Pat Hanrahan}.} \bibinfo{year}{2023}\natexlab{}.
\newblock \showarticletitle{Light field rendering}.
\newblock In \bibinfo{booktitle}{\emph{Seminal Graphics Papers: Pushing the Boundaries, Volume 2}}. \bibinfo{pages}{441--452}.
\newblock


\bibitem[Lin et~al\mbox{.}(2018)]%
        {lin2018learning}
\bibfield{author}{\bibinfo{person}{Chen-Hsuan Lin}, \bibinfo{person}{Chen Kong}, {and} \bibinfo{person}{Simon Lucey}.} \bibinfo{year}{2018}\natexlab{}.
\newblock \showarticletitle{Learning efficient point cloud generation for dense 3d object reconstruction}. In \bibinfo{booktitle}{\emph{proceedings of the AAAI Conference on Artificial Intelligence}}, Vol.~\bibinfo{volume}{32}.
\newblock


\bibitem[Lindell et~al\mbox{.}(2021)]%
        {lindell2021autoint}
\bibfield{author}{\bibinfo{person}{David~B Lindell}, \bibinfo{person}{Julien~NP Martel}, {and} \bibinfo{person}{Gordon Wetzstein}.} \bibinfo{year}{2021}\natexlab{}.
\newblock \showarticletitle{Autoint: Automatic integration for fast neural volume rendering}. In \bibinfo{booktitle}{\emph{Proceedings of the IEEE/CVF Conference on Computer Vision and Pattern Recognition}}. \bibinfo{pages}{14556--14565}.
\newblock


\bibitem[Liu et~al\mbox{.}(2020)]%
        {liu2020neural}
\bibfield{author}{\bibinfo{person}{Lingjie Liu}, \bibinfo{person}{Jiatao Gu}, \bibinfo{person}{Kyaw Zaw~Lin}, \bibinfo{person}{Tat-Seng Chua}, {and} \bibinfo{person}{Christian Theobalt}.} \bibinfo{year}{2020}\natexlab{}.
\newblock \showarticletitle{Neural sparse voxel fields}.
\newblock \bibinfo{journal}{\emph{Advances in Neural Information Processing Systems}}  \bibinfo{volume}{33} (\bibinfo{year}{2020}), \bibinfo{pages}{15651--15663}.
\newblock


\bibitem[Lombardi et~al\mbox{.}(2021)]%
        {lombardi2021mixture}
\bibfield{author}{\bibinfo{person}{Stephen Lombardi}, \bibinfo{person}{Tomas Simon}, \bibinfo{person}{Gabriel Schwartz}, \bibinfo{person}{Michael Zollhoefer}, \bibinfo{person}{Yaser Sheikh}, {and} \bibinfo{person}{Jason Saragih}.} \bibinfo{year}{2021}\natexlab{}.
\newblock \showarticletitle{Mixture of volumetric primitives for efficient neural rendering}.
\newblock \bibinfo{journal}{\emph{ACM Transactions on Graphics (ToG)}} \bibinfo{volume}{40}, \bibinfo{number}{4} (\bibinfo{year}{2021}), \bibinfo{pages}{1--13}.
\newblock


\bibitem[Lu et~al\mbox{.}(2023)]%
        {scaffoldgs}
\bibfield{author}{\bibinfo{person}{Tao Lu}, \bibinfo{person}{Mulin Yu}, \bibinfo{person}{Linning Xu}, \bibinfo{person}{Yuanbo Xiangli}, \bibinfo{person}{Limin Wang}, \bibinfo{person}{Dahua Lin}, {and} \bibinfo{person}{Bo Dai}.} \bibinfo{year}{2023}\natexlab{}.
\newblock \bibinfo{title}{Scaffold-GS: Structured 3D Gaussians for View-Adaptive Rendering}.
\newblock
\newblock
\showeprint[arxiv]{2312.00109}~[cs.CV]


\bibitem[Luiten et~al\mbox{.}(2023)]%
        {luiten2023dynamic}
\bibfield{author}{\bibinfo{person}{Jonathon Luiten}, \bibinfo{person}{Georgios Kopanas}, \bibinfo{person}{Bastian Leibe}, {and} \bibinfo{person}{Deva Ramanan}.} \bibinfo{year}{2023}\natexlab{}.
\newblock \showarticletitle{Dynamic 3d gaussians: Tracking by persistent dynamic view synthesis}.
\newblock \bibinfo{journal}{\emph{arXiv preprint arXiv:2308.09713}} (\bibinfo{year}{2023}).
\newblock


\bibitem[Martin-Brualla et~al\mbox{.}(2021)]%
        {nerf-in-the-wild}
\bibfield{author}{\bibinfo{person}{Ricardo Martin-Brualla}, \bibinfo{person}{Noha Radwan}, \bibinfo{person}{Mehdi~SM Sajjadi}, \bibinfo{person}{Jonathan~T Barron}, \bibinfo{person}{Alexey Dosovitskiy}, {and} \bibinfo{person}{Daniel Duckworth}.} \bibinfo{year}{2021}\natexlab{}.
\newblock \showarticletitle{Nerf in the wild: Neural radiance fields for unconstrained photo collections}. In \bibinfo{booktitle}{\emph{Proceedings of the IEEE/CVF Conference on Computer Vision and Pattern Recognition}}. \bibinfo{pages}{7210--7219}.
\newblock


\bibitem[Mildenhall et~al\mbox{.}(2020)]%
        {nerf}
\bibfield{author}{\bibinfo{person}{Ben Mildenhall}, \bibinfo{person}{Pratul~P. Srinivasan}, \bibinfo{person}{Matthew Tancik}, \bibinfo{person}{Jonathan~T. Barron}, \bibinfo{person}{Ravi Ramamoorthi}, {and} \bibinfo{person}{Ren Ng}.} \bibinfo{year}{2020}\natexlab{}.
\newblock \showarticletitle{NeRF: Representing Scenes as Neural Radiance Fields for View Synthesis}. In \bibinfo{booktitle}{\emph{ECCV}}.
\newblock


\bibitem[M{\"u}ller et~al\mbox{.}(2022a)]%
        {muller2022instant}
\bibfield{author}{\bibinfo{person}{Thomas M{\"u}ller}, \bibinfo{person}{Alex Evans}, \bibinfo{person}{Christoph Schied}, {and} \bibinfo{person}{Alexander Keller}.} \bibinfo{year}{2022}\natexlab{a}.
\newblock \showarticletitle{Instant neural graphics primitives with a multiresolution hash encoding}.
\newblock \bibinfo{journal}{\emph{ACM Transactions on Graphics (ToG)}} \bibinfo{volume}{41}, \bibinfo{number}{4} (\bibinfo{year}{2022}), \bibinfo{pages}{1--15}.
\newblock


\bibitem[M{\"u}ller et~al\mbox{.}(2022b)]%
        {ngp}
\bibfield{author}{\bibinfo{person}{Thomas M{\"u}ller}, \bibinfo{person}{Alex Evans}, \bibinfo{person}{Christoph Schied}, {and} \bibinfo{person}{Alexander Keller}.} \bibinfo{year}{2022}\natexlab{b}.
\newblock \showarticletitle{Instant neural graphics primitives with a multiresolution hash encoding}.
\newblock \bibinfo{journal}{\emph{ACM Transactions on Graphics (ToG)}} \bibinfo{volume}{41}, \bibinfo{number}{4} (\bibinfo{year}{2022}), \bibinfo{pages}{1--15}.
\newblock


\bibitem[Neff et~al\mbox{.}(2021)]%
        {neff2021donerf}
\bibfield{author}{\bibinfo{person}{Thomas Neff}, \bibinfo{person}{Pascal Stadlbauer}, \bibinfo{person}{Mathias Parger}, \bibinfo{person}{Andreas Kurz}, \bibinfo{person}{Joerg~H Mueller}, \bibinfo{person}{Chakravarty R~Alla Chaitanya}, \bibinfo{person}{Anton Kaplanyan}, {and} \bibinfo{person}{Markus Steinberger}.} \bibinfo{year}{2021}\natexlab{}.
\newblock \showarticletitle{DONeRF: Towards Real-Time Rendering of Compact Neural Radiance Fields using Depth Oracle Networks}. In \bibinfo{booktitle}{\emph{Computer Graphics Forum}}, Vol.~\bibinfo{volume}{40}. Wiley Online Library, \bibinfo{pages}{45--59}.
\newblock


\bibitem[Penner and Zhang(2017)]%
        {penner2017soft}
\bibfield{author}{\bibinfo{person}{Eric Penner} {and} \bibinfo{person}{Li Zhang}.} \bibinfo{year}{2017}\natexlab{}.
\newblock \showarticletitle{Soft 3d reconstruction for view synthesis}.
\newblock \bibinfo{journal}{\emph{ACM Transactions on Graphics (TOG)}} \bibinfo{volume}{36}, \bibinfo{number}{6} (\bibinfo{year}{2017}), \bibinfo{pages}{1--11}.
\newblock


\bibitem[Qi et~al\mbox{.}(2016)]%
        {pointnet}
\bibfield{author}{\bibinfo{person}{Charles~R Qi}, \bibinfo{person}{Hao Su}, \bibinfo{person}{Kaichun Mo}, {and} \bibinfo{person}{Leonidas~J Guibas}.} \bibinfo{year}{2016}\natexlab{}.
\newblock \showarticletitle{PointNet: Deep Learning on Point Sets for 3D Classification and Segmentation}.
\newblock \bibinfo{journal}{\emph{arXiv preprint arXiv:1612.00593}} (\bibinfo{year}{2016}).
\newblock


\bibitem[Rakhimov et~al\mbox{.}(2022)]%
        {Rakhimov_2022_CVPR}
\bibfield{author}{\bibinfo{person}{Ruslan Rakhimov}, \bibinfo{person}{Andrei-Timotei Ardelean}, \bibinfo{person}{Victor Lempitsky}, {and} \bibinfo{person}{Evgeny Burnaev}.} \bibinfo{year}{2022}\natexlab{}.
\newblock \showarticletitle{NPBG++: Accelerating Neural Point-Based Graphics}. In \bibinfo{booktitle}{\emph{Proceedings of the IEEE/CVF Conference on Computer Vision and Pattern Recognition (CVPR)}}. \bibinfo{pages}{15969--15979}.
\newblock


\bibitem[Reiser et~al\mbox{.}(2021)]%
        {reiser2021kilonerf}
\bibfield{author}{\bibinfo{person}{Christian Reiser}, \bibinfo{person}{Songyou Peng}, \bibinfo{person}{Yiyi Liao}, {and} \bibinfo{person}{Andreas Geiger}.} \bibinfo{year}{2021}\natexlab{}.
\newblock \showarticletitle{Kilonerf: Speeding up neural radiance fields with thousands of tiny mlps}. In \bibinfo{booktitle}{\emph{Proceedings of the IEEE/CVF International Conference on Computer Vision}}. \bibinfo{pages}{14335--14345}.
\newblock


\bibitem[Ren et~al\mbox{.}(2002)]%
        {ren2002object}
\bibfield{author}{\bibinfo{person}{Liu Ren}, \bibinfo{person}{Hanspeter Pfister}, {and} \bibinfo{person}{Matthias Zwicker}.} \bibinfo{year}{2002}\natexlab{}.
\newblock \showarticletitle{Object space EWA surface splatting: A hardware accelerated approach to high quality point rendering}. In \bibinfo{booktitle}{\emph{Computer Graphics Forum}}, Vol.~\bibinfo{volume}{21}. Wiley Online Library, \bibinfo{pages}{461--470}.
\newblock


\bibitem[Riegler and Koltun(2021)]%
        {riegler2021stable}
\bibfield{author}{\bibinfo{person}{Gernot Riegler} {and} \bibinfo{person}{Vladlen Koltun}.} \bibinfo{year}{2021}\natexlab{}.
\newblock \showarticletitle{Stable view synthesis}. In \bibinfo{booktitle}{\emph{Proceedings of the IEEE/CVF Conference on Computer Vision and Pattern Recognition}}. \bibinfo{pages}{12216--12225}.
\newblock


\bibitem[{Sara Fridovich-Keil and Alex Yu} et~al\mbox{.}(2022)]%
        {plenoxels}
\bibfield{author}{\bibinfo{person}{{Sara Fridovich-Keil and Alex Yu}}, \bibinfo{person}{Matthew Tancik}, \bibinfo{person}{Qinhong Chen}, \bibinfo{person}{Benjamin Recht}, {and} \bibinfo{person}{Angjoo Kanazawa}.} \bibinfo{year}{2022}\natexlab{}.
\newblock \showarticletitle{Plenoxels: Radiance Fields without Neural Networks}. In \bibinfo{booktitle}{\emph{CVPR}}.
\newblock


\bibitem[Schonberger and Frahm(2016)]%
        {sfm}
\bibfield{author}{\bibinfo{person}{Johannes~L Schonberger} {and} \bibinfo{person}{Jan-Michael Frahm}.} \bibinfo{year}{2016}\natexlab{}.
\newblock \showarticletitle{Structure-from-motion revisited}. In \bibinfo{booktitle}{\emph{Proceedings of the IEEE conference on computer vision and pattern recognition}}. \bibinfo{pages}{4104--4113}.
\newblock


\bibitem[Sch\"{o}nberger and Frahm(2016)]%
        {colmap}
\bibfield{author}{\bibinfo{person}{Johannes~Lutz Sch\"{o}nberger} {and} \bibinfo{person}{Jan-Michael Frahm}.} \bibinfo{year}{2016}\natexlab{}.
\newblock \showarticletitle{{Structure-from-Motion Revisited}}. In \bibinfo{booktitle}{\emph{Conference on Computer Vision and Pattern Recognition (CVPR)}}.
\newblock


\bibitem[Srinivasan et~al\mbox{.}(2019)]%
        {srinivasan2019pushing}
\bibfield{author}{\bibinfo{person}{Pratul~P Srinivasan}, \bibinfo{person}{Richard Tucker}, \bibinfo{person}{Jonathan~T Barron}, \bibinfo{person}{Ravi Ramamoorthi}, \bibinfo{person}{Ren Ng}, {and} \bibinfo{person}{Noah Snavely}.} \bibinfo{year}{2019}\natexlab{}.
\newblock \showarticletitle{Pushing the boundaries of view extrapolation with multiplane images}. In \bibinfo{booktitle}{\emph{Proceedings of the IEEE/CVF Conference on Computer Vision and Pattern Recognition}}. \bibinfo{pages}{175--184}.
\newblock


\bibitem[Vaswani et~al\mbox{.}(2017)]%
        {attn}
\bibfield{author}{\bibinfo{person}{Ashish Vaswani}, \bibinfo{person}{Noam Shazeer}, \bibinfo{person}{Niki Parmar}, \bibinfo{person}{Jakob Uszkoreit}, \bibinfo{person}{Llion Jones}, \bibinfo{person}{Aidan~N Gomez}, \bibinfo{person}{{\L}ukasz Kaiser}, {and} \bibinfo{person}{Illia Polosukhin}.} \bibinfo{year}{2017}\natexlab{}.
\newblock \showarticletitle{Attention is all you need}.
\newblock \bibinfo{journal}{\emph{Advances in neural information processing systems}}  \bibinfo{volume}{30} (\bibinfo{year}{2017}).
\newblock


\bibitem[Wang et~al\mbox{.}(2022)]%
        {r2l}
\bibfield{author}{\bibinfo{person}{Huan Wang}, \bibinfo{person}{Jian Ren}, \bibinfo{person}{Zeng Huang}, \bibinfo{person}{Kyle Olszewski}, \bibinfo{person}{Menglei Chai}, \bibinfo{person}{Yun Fu}, {and} \bibinfo{person}{Sergey Tulyakov}.} \bibinfo{year}{2022}\natexlab{}.
\newblock \showarticletitle{R2l: Distilling neural radiance field to neural light field for efficient novel view synthesis}. In \bibinfo{booktitle}{\emph{European Conference on Computer Vision}}. Springer, \bibinfo{pages}{612--629}.
\newblock


\bibitem[Wang et~al\mbox{.}(2004)]%
        {ssim}
\bibfield{author}{\bibinfo{person}{Zhou Wang}, \bibinfo{person}{Alan~C Bovik}, \bibinfo{person}{Hamid~R Sheikh}, {and} \bibinfo{person}{Eero~P Simoncelli}.} \bibinfo{year}{2004}\natexlab{}.
\newblock \showarticletitle{Image quality assessment: from error visibility to structural similarity}.
\newblock \bibinfo{journal}{\emph{IEEE transactions on image processing}} \bibinfo{volume}{13}, \bibinfo{number}{4} (\bibinfo{year}{2004}), \bibinfo{pages}{600--612}.
\newblock


\bibitem[Wiles et~al\mbox{.}(2020)]%
        {wiles2020synsin}
\bibfield{author}{\bibinfo{person}{Olivia Wiles}, \bibinfo{person}{Georgia Gkioxari}, \bibinfo{person}{Richard Szeliski}, {and} \bibinfo{person}{Justin Johnson}.} \bibinfo{year}{2020}\natexlab{}.
\newblock \showarticletitle{Synsin: End-to-end view synthesis from a single image}. In \bibinfo{booktitle}{\emph{Proceedings of the IEEE/CVF Conference on Computer Vision and Pattern Recognition}}. \bibinfo{pages}{7467--7477}.
\newblock


\bibitem[Wu et~al\mbox{.}(2023)]%
        {wu20234d}
\bibfield{author}{\bibinfo{person}{Guanjun Wu}, \bibinfo{person}{Taoran Yi}, \bibinfo{person}{Jiemin Fang}, \bibinfo{person}{Lingxi Xie}, \bibinfo{person}{Xiaopeng Zhang}, \bibinfo{person}{Wei Wei}, \bibinfo{person}{Wenyu Liu}, \bibinfo{person}{Qi Tian}, {and} \bibinfo{person}{Xinggang Wang}.} \bibinfo{year}{2023}\natexlab{}.
\newblock \showarticletitle{4d gaussian splatting for real-time dynamic scene rendering}.
\newblock \bibinfo{journal}{\emph{arXiv preprint arXiv:2310.08528}} (\bibinfo{year}{2023}).
\newblock


\bibitem[Yang et~al\mbox{.}(2023)]%
        {yang2023deformable}
\bibfield{author}{\bibinfo{person}{Ziyi Yang}, \bibinfo{person}{Xinyu Gao}, \bibinfo{person}{Wen Zhou}, \bibinfo{person}{Shaohui Jiao}, \bibinfo{person}{Yuqing Zhang}, {and} \bibinfo{person}{Xiaogang Jin}.} \bibinfo{year}{2023}\natexlab{}.
\newblock \showarticletitle{Deformable 3d gaussians for high-fidelity monocular dynamic scene reconstruction}.
\newblock \bibinfo{journal}{\emph{arXiv preprint arXiv:2309.13101}} (\bibinfo{year}{2023}).
\newblock


\bibitem[Yifan et~al\mbox{.}(2019)]%
        {yifan2019differentiable}
\bibfield{author}{\bibinfo{person}{Wang Yifan}, \bibinfo{person}{Felice Serena}, \bibinfo{person}{Shihao Wu}, \bibinfo{person}{Cengiz {\"O}ztireli}, {and} \bibinfo{person}{Olga Sorkine-Hornung}.} \bibinfo{year}{2019}\natexlab{}.
\newblock \showarticletitle{Differentiable surface splatting for point-based geometry processing}.
\newblock \bibinfo{journal}{\emph{ACM Transactions on Graphics (TOG)}} \bibinfo{volume}{38}, \bibinfo{number}{6} (\bibinfo{year}{2019}), \bibinfo{pages}{1--14}.
\newblock


\bibitem[Yu et~al\mbox{.}(2021)]%
        {yu2021plenoctrees}
\bibfield{author}{\bibinfo{person}{Alex Yu}, \bibinfo{person}{Ruilong Li}, \bibinfo{person}{Matthew Tancik}, \bibinfo{person}{Hao Li}, \bibinfo{person}{Ren Ng}, {and} \bibinfo{person}{Angjoo Kanazawa}.} \bibinfo{year}{2021}\natexlab{}.
\newblock \showarticletitle{Plenoctrees for real-time rendering of neural radiance fields}. In \bibinfo{booktitle}{\emph{Proceedings of the IEEE/CVF International Conference on Computer Vision}}. \bibinfo{pages}{5752--5761}.
\newblock


\bibitem[Zhang et~al\mbox{.}(2018)]%
        {lpips}
\bibfield{author}{\bibinfo{person}{R. Zhang}, \bibinfo{person}{P. Isola}, \bibinfo{person}{A.~A. Efros}, \bibinfo{person}{E. Shechtman}, {and} \bibinfo{person}{O. Wang}.} \bibinfo{year}{2018}\natexlab{}.
\newblock \showarticletitle{The Unreasonable Effectiveness of Deep Features as a Perceptual Metric}. In \bibinfo{booktitle}{\emph{2018 IEEE/CVF Conference on Computer Vision and Pattern Recognition (CVPR)}}. \bibinfo{pages}{586--595}.
\newblock


\bibitem[Zhou et~al\mbox{.}(2018)]%
        {zhou2018stereo}
\bibfield{author}{\bibinfo{person}{Tinghui Zhou}, \bibinfo{person}{Richard Tucker}, \bibinfo{person}{John Flynn}, \bibinfo{person}{Graham Fyffe}, {and} \bibinfo{person}{Noah Snavely}.} \bibinfo{year}{2018}\natexlab{}.
\newblock \showarticletitle{Stereo magnification: Learning view synthesis using multiplane images}.
\newblock \bibinfo{journal}{\emph{arXiv preprint arXiv:1805.09817}} (\bibinfo{year}{2018}).
\newblock


\bibitem[Zwicker et~al\mbox{.}(2001)]%
        {ewa}
\bibfield{author}{\bibinfo{person}{Matthias Zwicker}, \bibinfo{person}{Hanspeter Pfister}, \bibinfo{person}{Jeroen Van~Baar}, {and} \bibinfo{person}{Markus Gross}.} \bibinfo{year}{2001}\natexlab{}.
\newblock \showarticletitle{EWA volume splatting}. In \bibinfo{booktitle}{\emph{Proceedings Visualization, 2001. VIS'01.}} IEEE, \bibinfo{pages}{29--538}.
\newblock


\end{thebibliography}
